\documentclass[11pt]{article}
\usepackage[utf8]{inputenc}
\usepackage{amssymb,verbatim,epsfig}
\usepackage{amsmath,setspace}
\usepackage{siunitx}
\usepackage{multicol}
\usepackage{multirow}
\usepackage{vcell}
\usepackage{url}
\usepackage[dvipsnames]{xcolor}
\usepackage{wrapfig}
\usepackage[capitalize]{cleveref}
\usepackage{amssymb, amsmath, verbatim, epsfig}
\usepackage{setspace}
\usepackage{longtable}
\usepackage{float}
\usepackage{lineno}
\usepackage[margin=1in]{geometry} 
\usepackage[small,bf]{caption}
\usepackage[sort&compress,comma,super]{natbib}
\usepackage{enumitem}

\newcommand{\eq}[1]{Eq.(\ref{#1})}
\newcommand{\fig}[1]{{\bf Fig.\ref{#1}}}
\newcommand{\tab}[1]{{\bf Table \ref{#1}}}
\newcommand{\et}{{\it et al.~}}

\newcommand{\volume}{{\ooalign{\hfil$V$\hfil\cr\kern0.08em--\hfil\cr}}}
\spacing{1}

\title{\bf 
\vspace{-0.75in}
Cavitation-induced microjets tuned by channels with alternating wettability patterns} 

\author{ Jelle J. Schoppink$^{1\dag}$, Keerthana Mohan$^{1\dag}$, Miguel A.~Quetzeri-Santiago$^1$, \\ Gareth McKinley$^2$, 
David Fernandez Rivas$^1$, \& Andrew K.~Dickerson$^{3*}$\\ 
\small $^1$Mesoscale Chemical Systems Group, Univ. of Twente, Enschede, Netherlands\\
\small $^2$ Mechanical Engineering, Massachusetts Institute of Technology, Cambridge, MA 02139 \\
  \small  $^3$Mechanical, Aerospace, and Biomedical Eng., Univ. of Tennessee, Knoxville, TN 37996\\
  \small $^\dag$ These two authors contributed equally\\
   \small $^*$ Corresponding author: ad@utk.edu}
\date{}
\begin{document} \maketitle 
\begin{abstract}
A laser pulse focused near the closed end of a glass capillary partially filled with water creates a vapor bubble and an associated pressure wave. The pressure wave travels through the liquid toward the meniscus where it is reflected, creating a fast, focused microjet. In this study, we selectively coat the hydrophilic glass capillaries with hydrophobic strips along the capillary. The result after filling the capillary is a static meniscus which has a curvature markedly different than an unmodified capillary. This tilting asymmetry in the static meniscus alters the trajectory of the ensuing jets. The hydrophobic strips also influence the advancing contact line and receding contact line as the vapor bubble expands and collapses. We present thirteen different permutations of this system which includes three geometries and four coating schemes. The combination of geometry and coatings influences the jet breakup, the resulting drop size distribution, the trajectory of the jet tip, and the consistency of jet characteristics across trials. The inclusion of hydrophobic strips promotes jetting in line with the channel axis, with the most effective arrangement dependent on channel size. 
\end{abstract}

\maketitle
\vspace{0.1 in}
\noindent{\bf Keywords:} []
\section{Introduction}\label{intro}

The stability of liquid jets has captivated fluid mechanicians for nearly two centuries \cite{bidone1829experiences, savart1833notes}, owing to both their mathematical complexity \cite{amini2014instability, ashgriz2011capillary, eggers2008physics, sirignano2000review, entov1984dynamics, entov1980dynamical} and usefulness \cite{jarrahbashi2012acceleration,mitragotri2006current,reneker2000bending}. Recently, jet dynamics have gained attention from the engineering and medical communities for their use in drug delivery \cite{menezes2009shock, tagawa, mitragotri2006current}, ink-jet printers \cite{furlani2006thermally, castrejon2013future}, and micro-fabrication \cite{carter2006fabricating, macfarlane1994microjet}. Such microscale jets rely on the sudden acceleration of a liquid column\cite{kiyama2016effects,antkowiak2007short}, piezoelectric actuation \cite{reis2005ink} or by rapidly vaporizing a portion of liquid upstream with a laser pulse (thermocavitation)\cite{oyarte2020microfluidics, tagawa2012highly,peters2013highly,Schoppink2022perspective}.
Impulsively created jets are unsteady and are ``kinematically focused" by a curved meniscus in which a pressure wave is reflected at the free surface\cite{tagawa2012highly}. The focused liquid converges toward the center of curvature, resulting in jets with velocities that can exceed a Mach number\cite{tagawa2012highly,  Krizek2020SciRep} $M=U/c_\text{s}=1$, where $U$ is the jet velocity and $c_\text{s}$ is the speed of sound in air. Jets emerge from the focused menisci in the form of a stretched ligament that breaks into droplets. Numerous theoretical and experimental investigations have been carried out to explain the disintegration of liquid jets, which are inherently unstable \cite{eggers2008physics}. Various forces act on the surface of jets leading to disturbances that are amplified when carried downstream, ultimately leading to jet breakup by the Rayleigh-Plateau instability among others \cite{Rayleigh_jet, Ashgriz:2011aa, eggers2008physics}.

The number and trajectory of droplets after jet breakup are guided by the characteristics of the initial impulse, bubble retraction in the case of thermocavitation, meniscus shape\cite{rodriguez2017toward, gordillo2020impulsive}, and contact line motion. When the jet leaves the nozzle, the no-slip boundary condition is relieved at the outer radial edge, leading to the creation of radial velocity components within the jet. This profile relaxation generates instabilities in the jet \cite{sterling1975instability}. Therefore, the jet characteristics can be modulated by modifying the initial meniscus shape and the nature of contact line motion. 

The orifice geometry is another variable influencing jet disintegration \cite{Birouk_2009,wang2015liquid}. For non-circular nozzles, the propagating jet expands along one radial axis, while contracting in the other in an oscillating manner, destabilizing the jet\cite{wang2015liquid}. This so-called axis switching has been modeled as a spring-mass system driven by the competition of surface tension and inertia \cite{amini2012axis}. Jets produced by non-circular nozzles break up into smaller droplets and have shorter breakup lengths than comparable circular nozzles \cite{wang2015liquid}.  These chain-like oscillations in the jet are caused by non-axisymmetric perturbations which are less unstable than Rayleigh-Plateau instabilities.  Chain-like oscillations are non-linear in nature and their frequency decreases with increasing amplitude \cite{Jordan2022Chain}. In a typical jetting experiment, the Rayleigh-Plateau instability is superimposed on non-axisymmetric perturbations to cause jet breakup \cite{Jordan2022Chain}.  

\begin{figure}[h]
    \centering
    \includegraphics[width=0.75\textwidth]{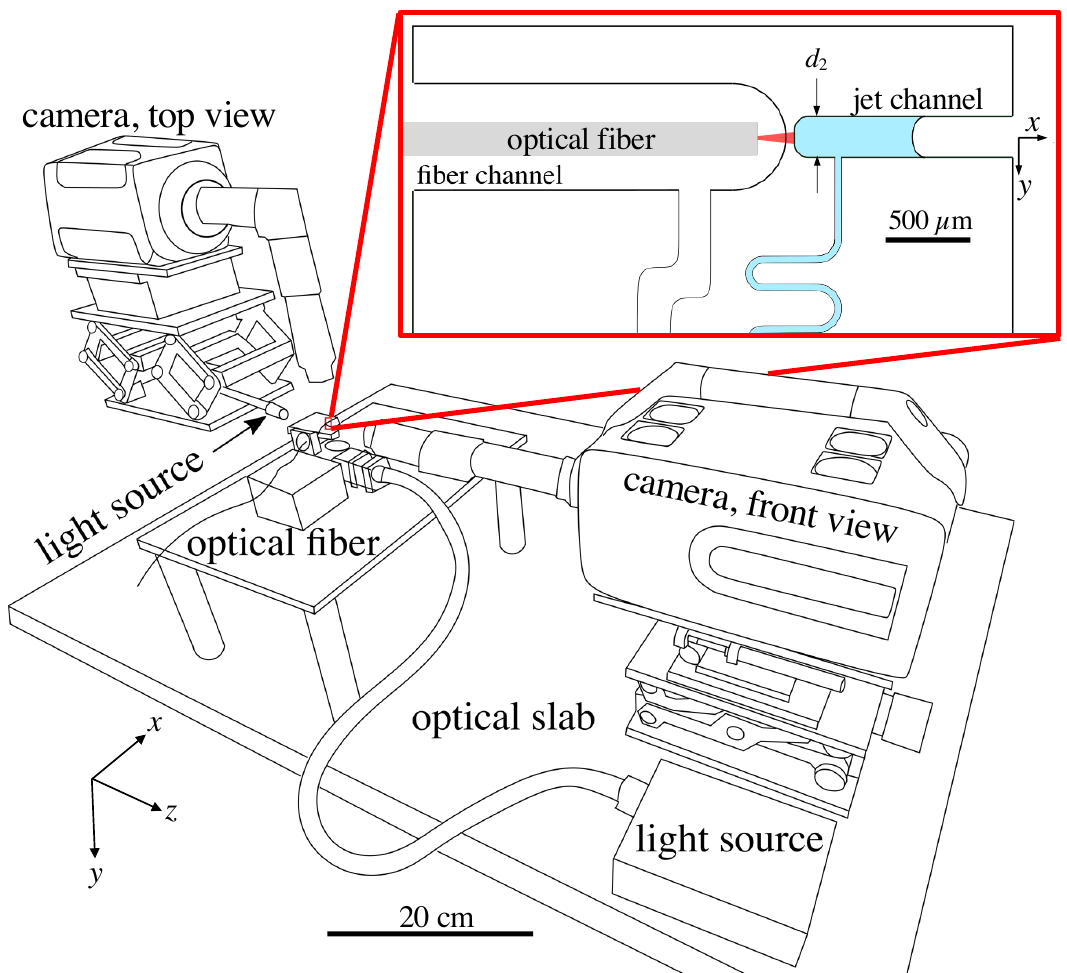}
    \caption{Schematic of the experimental setup showing the orientation of the chip with respect to both camera views. The zoom box shows a glass chip in detail. Jets emerge from chips to the right.}
    \label{setup}
\end{figure}

\begin{figure}[h]
    \centering
    \includegraphics[width=\textwidth]{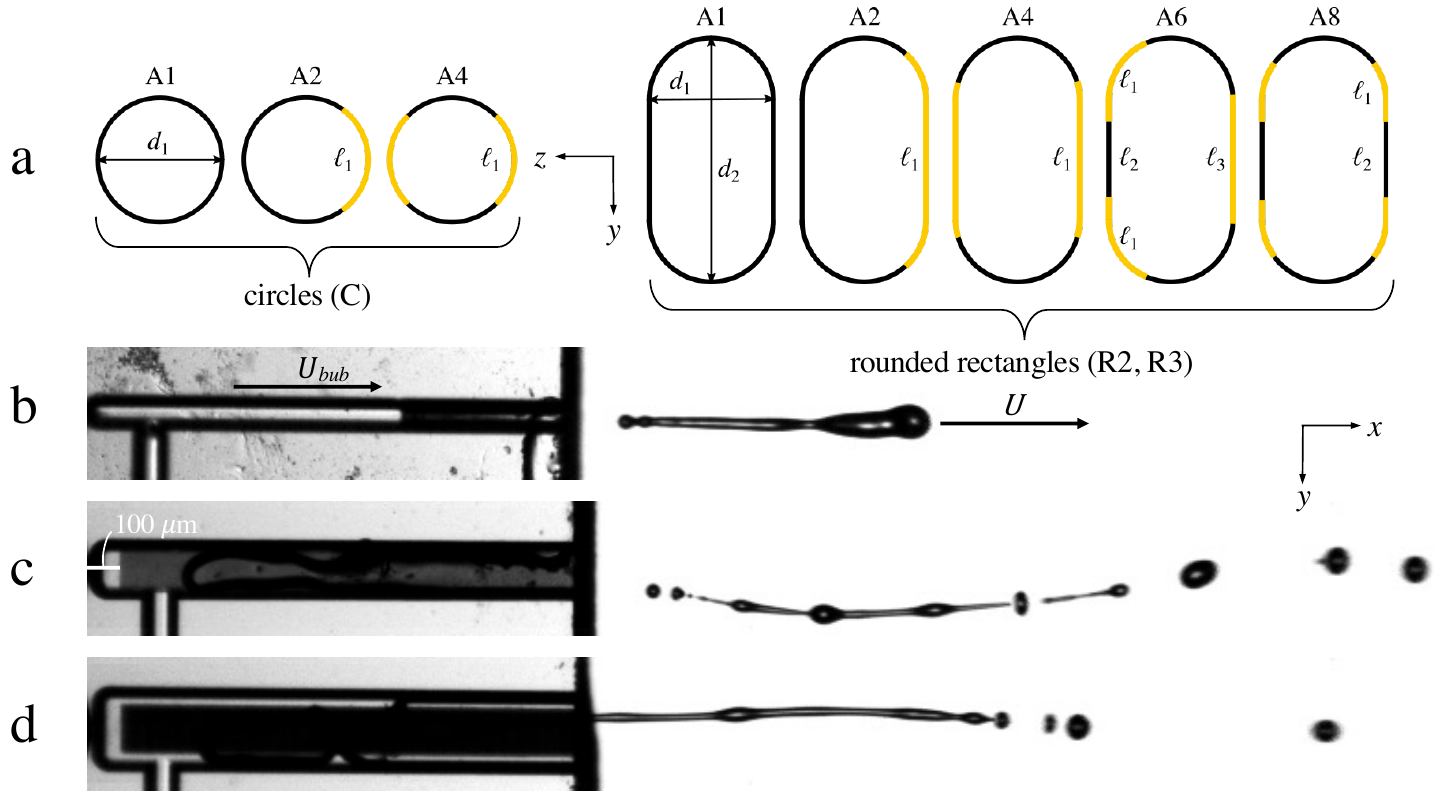}
    \caption{\textbf{(a)} Schematic of channel inner surface coating and cross-section permutations, and representative jet images from channels. Diagrams are such that the camera views them looking from left to right.  \textbf{(b)} CA1, \textbf{(c)} R2A2, and \textbf{(d)} R3A4. Every channel is 1850 $\mu$m long and 100 $\mu$m deep into the frame. Other pertinent dimensions are given in \tab{dims_tab}.}
    \label{Intro_fig}
\end{figure}

In this study we use an infrared laser pulse to create a cavitation bubble at the closed end of a microscale liquid channel and film the expulsion of the jet from two perpendicular views, as shown in \fig{setup}. Our experimental system is similar to that established by Oyarte Gálvez \et (2020)\cite{oyarte2020microfluidics} and earlier by Tagawa \et (2012)\cite{tagawa2012highly}, but here we probe how the channel geometry and its wettability influence the jet characteristics. Two fundamental channel shapes are etched into borosilicate chips for experimental investigation, circular (C) and rounded rectangular (R) cross-sections, as shown in \fig{Intro_fig}a.  The full length and relative height of each tested geometry, and the variety of jets they produce, are shown in \fig{Intro_fig}b-d. 
Channel surface chemistry is either homogeneous (A1) or has alternating hydrophobic-hydrophilic sections (A2-A8) as depicted in \fig{Intro_fig}a. 
Due to manufacturing limitations, circular cross-sections have only three coating permutations, whereas rounded rectangles have five, for a total of thirteen unique channel configurations.

The jets created under the conditions tested in this study are of interest for the role they will play in needle-free injection and other microscale liquid delivery devices. However, our primary goal is  to unravel the connection between channel geometry and subsequent jet properties, which may be useful for other applications such as coating and spraying of surfaces.

The fabrication methods, experimental protocols, and further details of chip design are given in Section \ref{methods}. We present experimental results and discussion of jet velocity, droplet characteristics, repeatability, and focusing in Section \ref{results}, by splitting the observed phenomena into axial (Sections 3.1-3.4) and out-of-axis (Sections 3.5-3.9) behavior. We conclude our work in Section \ref{conclusion}.

 

  
 
 



\section{Experimental methods}\label{methods}
\subsection{Chip layout and fabrication}

The overall layout of our experimental microfluidic chips is shown in \fig{setup}. 
The fiber channel is nearly cylindrical and measures 425 $\mu$m tall, 400 $\mu$m deep, and 2450 $\mu$m long. A 400-$\mu$m channel is included to serve as an inlet and flush the fiber channel after fabrication to remove contaminants.
All jet channels have a characteristic cross-section width $d_1=100$ $\mu$m and are 1850 $\mu$m in length. Rounded rectangles have two size configurations, R2 and R3, such that $d_2/d_1 = 2$ and 3, respectively. Relative channel size is shown in \fig{Intro_fig}. The channel cross-sectional area $A$ and perimeter $P$ are computed using that of rectangles capped by two half-circles. Channel area $A$ and hydraulic diameter $D_\text{H}=4A/P$ are reported in \tab{dims_tab}.
Homogeneous glass channels (A1) are modified by selectively depositing atomically thin layers of gold that are thereafter soaked in thiol. The result is that channels have \textit{alternating} sections of hydrophilic glass, $\theta_\text{e}\approx 30^\circ$, and hydrophobic gold, $\theta_\text{e}\approx 115^\circ$, where $\theta_\text{e}$ is the equilibrium contact angle.
The arclength of coated and uncoated sections $\ell_1$, $\ell_2$, and $\ell_3$ are shown schematically in \fig{Intro_fig}a and provided in \tab{dims_tab}. 
We henceforth refer to channels by an abbreviated identifier. For example, a rounded rectangle with a cross-sectional aspect ratio of three and six discrete alternating sections is referred to by R3A6.
The jet channel is filled by a 360-$\mu$m glass capillary with distilled water, as shown in \fig{setup}. The circular fill channel is 100 $\mu$m in diameter and meanders to provide greater hydraulic resistance such that flow is preferential down the jet channel rather than towards the filling channel. Each gold strip begins 100 $\mu$m from the closed end of the jet channel, as shown in \fig{Intro_fig}c.

Glass chips are fabricated under cleanroom conditions. The channel structures with half depths of 50 and 200 $\mu$m are wet etched into 4 inch borosilicate glass wafers with a thickness of 500 $\mu$m. Next, a new photoresist is applied to the glass wafers and removed from the intended position of the gold structures. A 15-nm thick coating of tantalum is applied prior to a 45-nm thick coating of gold. The photoresist is removed, and a gold layer remains only on the intended positions. Afterward, two glass wafers are bonded together and diced to create single chips.
The gold surface is made hydrophobic according to Notsu \et (2005)\cite{notsu2005super}. The chips are immersed for one hour into a 10-mM solution of 1H,1H,2H,2H-perfluorodecanethiol (PFDT, Sigma Aldrich) in ethanol, after which the channels are briefly flushed with ethanol to remove any excess PFDT. 
The PFDT has no effect on the borosilicate glass.

\begin{longtable}{|c|ccccccccccccc|} 
\caption{Channel and coating parameters.} \label{dims_tab}
\\
\hline
\begin{tabular}[c]{@{}c@{}}Geometric\\ Configuration\end{tabular} & \multicolumn{3}{c|}{\textbf{C}}                                                                        & \multicolumn{5}{c|}{\textbf{R2}}                                                                                                                                             & \multicolumn{5}{c|}{\textbf{R3}}                                                                                                                        \\ \hline
\endfirsthead
\endhead
\begin{tabular}[c]{@{}c@{}}Coating\\ Configuration\end{tabular}   & \multicolumn{1}{c|}{\textbf{A1}} & \multicolumn{1}{c|}{\textbf{A2}} & \multicolumn{1}{c|}{\textbf{A4}} & \multicolumn{1}{c|}{\textbf{A1}} & \multicolumn{1}{c|}{\textbf{A2}} & \multicolumn{1}{c|}{\textbf{A4}} & \multicolumn{1}{c|}{\textbf{A6}} & \multicolumn{1}{c|}{\textbf{A8}} & \multicolumn{1}{c|}{\textbf{A1}} & \multicolumn{1}{c|}{\textbf{A2}} & \multicolumn{1}{c|}{\textbf{A4}} & \multicolumn{1}{c|}{\textbf{A6}} & \textbf{A8} \\ \hline
$d_1$ {[}$\mu$m{]}                                                & \multicolumn{13}{c|}{100}                                                                                                                                                                                                                                                                                                                                                                                                                       \\ \hline
$d_2$ {[}$\mu$m{]}                                                & \multicolumn{3}{c|}{--}                                                                                & \multicolumn{5}{c|}{200}                                                                                                                                                     & \multicolumn{5}{c|}{300}                                                                                                                                \\ \hline
$\ell_1$ {[}$\mu$m{]}                                             & \multicolumn{1}{c|}{--}          & \multicolumn{1}{c|}{103}         & \multicolumn{1}{c|}{84}          & \multicolumn{1}{c|}{--}          & \multicolumn{1}{c|}{193}         & \multicolumn{1}{c|}{128}         & \multicolumn{1}{c|}{77}          & \multicolumn{1}{c|}{64}          & \multicolumn{1}{c|}{--}          & \multicolumn{1}{c|}{293}         & \multicolumn{1}{c|}{178}         & \multicolumn{1}{c|}{97}          & 89          \\ \hline
$\ell_2$ {[}$\mu$m{]}                                             & \multicolumn{1}{c|}{--}          & \multicolumn{1}{c|}{--}          & \multicolumn{1}{c|}{--}          & \multicolumn{1}{c|}{--}          & \multicolumn{1}{c|}{--}          & \multicolumn{1}{c|}{--}          & \multicolumn{1}{c|}{64}          & \multicolumn{1}{c|}{64}          & \multicolumn{1}{c|}{--}          & \multicolumn{1}{c|}{--}          & \multicolumn{1}{c|}{--}          & \multicolumn{1}{c|}{98}          & 89          \\ \hline
$\ell_3$ {[}$\mu$m{]}                                             & \multicolumn{1}{c|}{--}          & \multicolumn{1}{c|}{--}          & \multicolumn{1}{c|}{--}          & \multicolumn{1}{c|}{--}          & \multicolumn{1}{c|}{--}          & \multicolumn{1}{c|}{--}          & \multicolumn{1}{c|}{107}         & \multicolumn{1}{c|}{--}          & \multicolumn{1}{c|}{--}          & \multicolumn{1}{c|}{--}          & \multicolumn{1}{c|}{--}          & \multicolumn{1}{c|}{107}         & --          \\ \hline

$A$ {[}$\mu$m$^2${]}                                                & \multicolumn{3}{c|}{7,854}                                                                                & \multicolumn{5}{c|}{17,854}                                                                                                                                                     & \multicolumn{5}{c|}{27,854}                                                                                                                                \\ \hline
$D_\text{H}$ {[}$\mu$m{]}                                                & \multicolumn{3}{c|}{100}                                                                                & \multicolumn{5}{c|}{139}                                                                                                                                                     & \multicolumn{5}{c|}{156}                                                                                                                                \\ \hline
\end{longtable} 

\subsection{Jet creation and high-speed imaging}\label{subsec: jet creation and filming}
Jets are created by the vaporization of water on the closed end of jetting channels by a 10~ms, 1.95~$\mu$m infrared laser pulse with a power of 0.59~$\pm$~0.03~W. 
Due to the high absorption coefficient of water at this wavelength ($\alpha$~$\approx$~$120$~cm$^{-1}$)~\cite{Ruru2012Absorption}, no dye is required, in contrast to our previous work~\cite{rodriguez2017toward, oyarte2020microfluidics, quetzeri2021impact}. The laser pulse is produced by a Thulium fiber laser (BKTel Photonics) with an SMF28 optical fiber output. The fiber is cleaved prior to use and the fiber tip is placed at a distance of 250 $\mu$m from the closed end of the jet channel. The laser power has a secondary fiber output of 1\% of the nominal power, which is monitored by a photodetector (Thorlabs DET05D2). To confirm the actual laser power, the photodetector is read out by an oscilloscope (Tektronix MSO 2014B).

Jetting events are filmed at two perpendicular angles by a Photron SA-X2 (Front view, $x,y$) and a Photron Nova S6 (Top view, $x,z$) at 144,000 fps. Both cameras are equipped with a Navitar 12$\times$ zoom lens, operating at magnifications of 3$\times$ and 2$\times$ respectively. Backlighting is provided by a Schott Coldvision-LS and SugarCUBE Ultra. All equipment is triggered simultaneously by an Arduino UNO.
Only select channel configurations were imaged from the top view. Priority was given to the asymmetrically coated channels. Therefore, the non-coated (CA1, R2A1, R3A1) and some symmetrically coated (R2A4, R2A8) are only imaged from the front. For the other eight chips, the jet was imaged from the front and top simultaneously.
Videos were processed by custom code in MATLAB.

\section{Results \& discussion}\label{results}

\begin{figure}
    \centering
    \includegraphics[width = 0.7\textwidth]{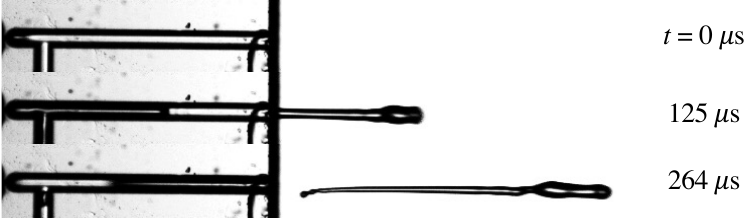}
    \caption{The jetting sequence of a fully filled CA1 channel. The panels show the moments of jet emergence from the channel (top), maximum bubble size (middle), and complete bubble collapse. (Multimedia View)}
    \label{S1}
\end{figure}
\begin{figure}[H]
    \centering
    \includegraphics[width=0.85\textwidth]{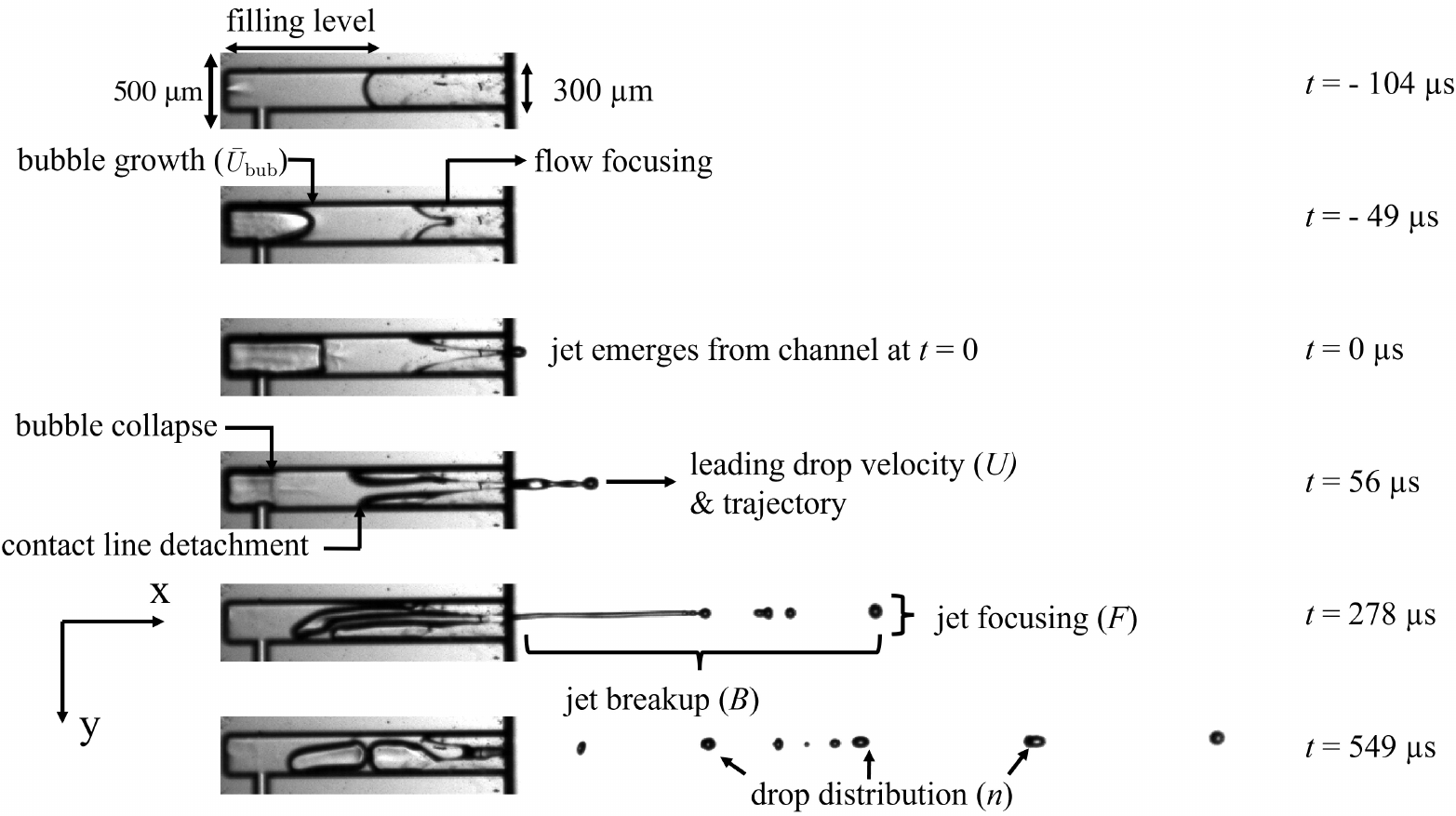}
    \caption{Sequence of images showing the jetting event. Each element of the complete jetting event will be discussed in a specific subsection: Bubble growth and initial jet formation, etc.}
    \label{introfig}
\end{figure}

We filmed approximately twenty jetting events from each thirteen channel configurations. The inclusion of PFDT-bonded gold coating in ten of the thirteen channels results in channels with a heterogeneous wetting condition. A typical jetting event is depicted in \fig{introfig}, where all the stages of the process are signaled. 
Channels are filled to half their length, $\sim 850~\mu$m, before activation of the laser pulse. Prior to bubble formation, a spot of non-homogeneous light intensity appears near the laser spot position, which we posit is due to an augmented refractive index caused by localized heating.
Expansion of the laser-induced bubble drives the kinematically focused meniscus forward. We set $t=0$ at the moment the jet emerges from the channel. 
A half-fill in our channels is done deliberately such that all moving menisci are allowed an equal and substantial runway length along the coated or uncoated channel walls. Channels that are fully filled do not experience the same degree of meniscus focusing and are only affected weakly by channel coating as the jet tip exits the channel, likely the result of not having a statically curved meniscus \cite{rodriguez2017toward}. An example of CA1 fully filled is provided in \fig{S1} (Multimedia View). Jets from completely filled channels tend to be larger in diameter, with a thicker tip \cite{rodriguez2017toward}. In the case of \fig{S1} (Multimedia View), the large tip flattens against air resistance as it emerges. On the other extreme, channels not sufficiently filled experience the breakup of the liquid plug before the jet exits, a phenomenon more likely as channel cross-sections grow in area from chip to chip. 


The sheer amount of data produced in this study precludes a full presentation of our results below, and we thus select six from the thirteen nozzles fabricated to feature in this main text. A comprehensive presentation of plots from all nozzle permutations is provided in Figs.S1-S13. For each geometric configuration, we feature the uncoated (A1) and a coated configuration providing the lowest focusing factor $F$, which is to be described in Section 3.6.
Our featured nozzles are labeled in the top left of the panels in \fig{bubs}.




\subsection{Bubble size and jet velocity}
The maximum bubble size, bubble expansion velocity, and filling level all influence jetting from a given channel geometry. Small bubbles (less than 20\% of channel length) create jets with little volume, often a few discrete droplets (of diameter ~50 $\mu m$), \fig{S2}.  However, all jets presented here are well above the transition from dripping to jetting\cite{Clanet}, given by a velocity at or less than 2.5 m/s. By contrast, larger bubbles 
produce a more complete emptying of the channel often with a curved jet tail (which we refer to as tail sway), see \fig{S3}. Bubbles with sizes comparable to the channel length can empty the channel almost completely, and the jet exits as a plug. The jet tip in this case is also thicker, similar to that of a fully filled channel. There exists an optimum range of bubble sizes versus filling levels for a given channel geometry in which a jet can be produced without the extreme cases of droplet or plug formation. For example, in the R2 channels, the transition to plug flow was observed for bubbles between 1.1 and 1.4 times the initial filling level, but its precise definition will be the topic of future work. 

\begin{figure}[H]
    \centering
    \includegraphics[width=0.85\textwidth]{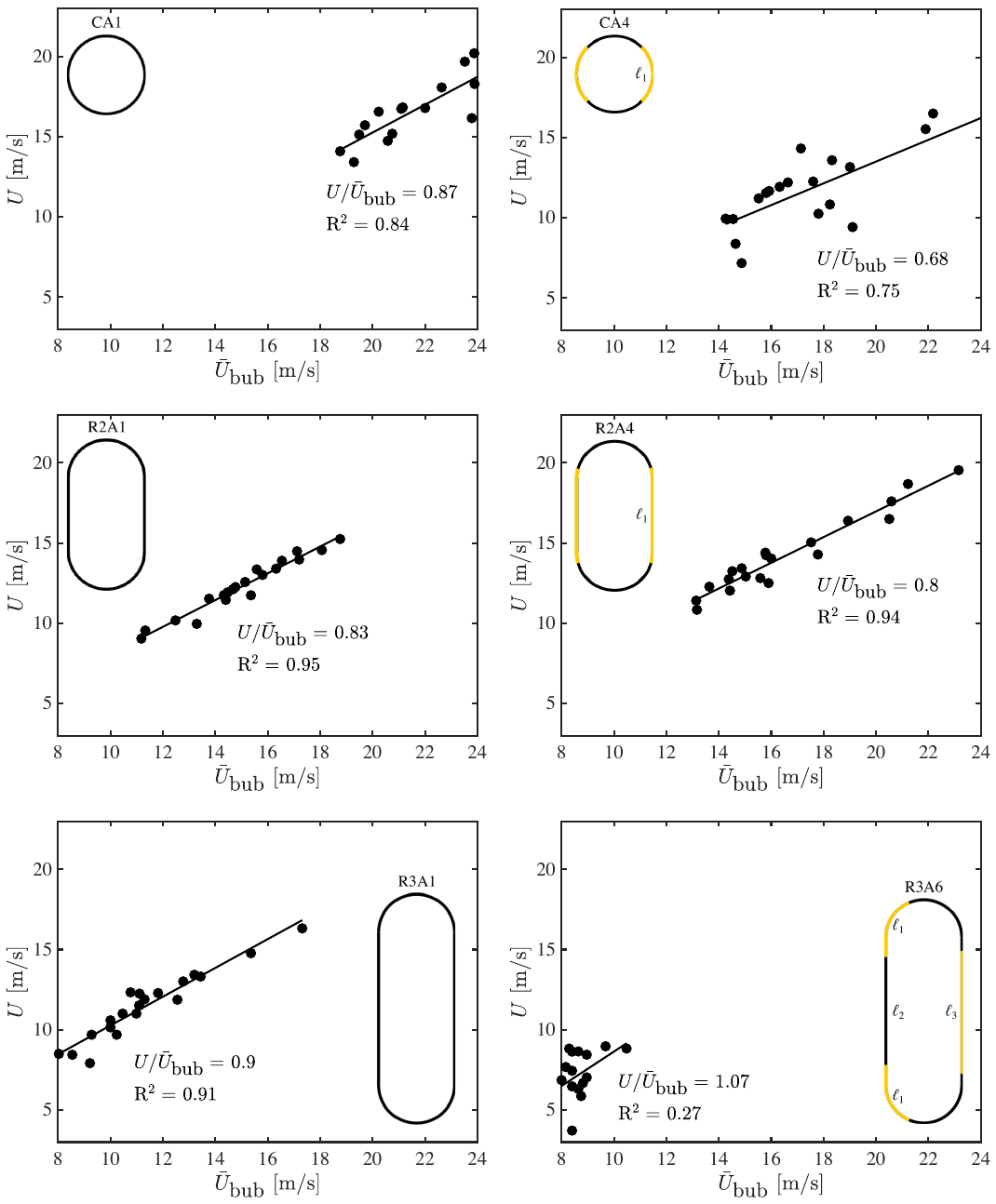}
    \caption{Jet velocity versus average bubble velocity for featured nozzles. For the circular channels, the hydrophobic coatings lower $U/\bar{U}_\mathrm{b}$ compared to uncoated channels. For R2 the effect of the coatings is negligible. In the case of larger R3 channels, coatings increase $U/\bar{U}_\mathrm{b}$ compared to uncoated channels.}
    \label{bubs}
\end{figure}

The relation between jet velocity ($U$) and the average velocity of the bubble front during its growth phase ($\bar{U}_\text{bub}$) is plotted in \fig{bubs}. We determine the jet velocity by tracking the leading edge of the jet tip from the moment it exits the chip at $t=0$ for ten frames (69.4 $\mu$s).  We here note the first distinction between jets from circular (C) and rectangular (R) channels. Both coated circular channels exhibit a lower aggregate $U/\bar{U}_\text{bub}$ when compared to CA1. In rectangular channel R3, coatings increase $U/\bar{U}_\text{bub}$ over the uncoated configuration. For all configurations $U/\bar{U}_\text{bub}$ and correlation coefficients for $U=j\bar{U}_\text{bub}$ are given in \tab{params_tab}, where $j$ is a fitting constant.

\begin{figure}
    \centering
    \includegraphics[width = 0.7\textwidth]{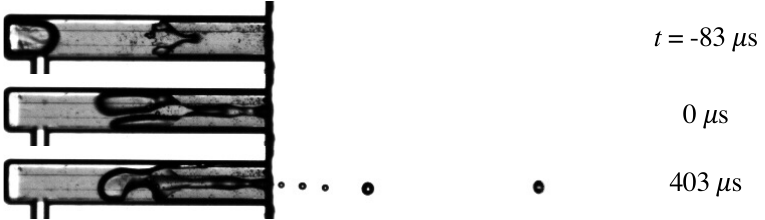}
    \caption{The jetting sequence produced by a relatively small bubble, less than 20\% of the channel length. The panels show the moments of maximum bubble size (top), jet emerge from the channel (middle), and when the last droplet breaks from the main jet body within the channel (bottom). (Multimedia View)}
    \label{S2}
\end{figure}

We use jet velocity $U$ and hydraulic diameter $D_\text{H}$ to define other common dimensionless groups used in jetting studies: Reynolds number $Re = \rho UD_\text{H}/\mu$, Weber number $We = \rho U^2 D_\text{H}/\sigma$, and Ohnesorge number $Oh = We/Re^2$. Here, the density, viscosity, and surface tension of distilled water are taken to be $\rho =1$ g/mL, $\mu =0.89$ cP, and $\sigma = 72.9$ dyne/cm, respectively. We report average $Re$ and $We$ for all channels in \tab{params_tab}. The range of Reynolds number, $175-3125$, indicates inertia dominates viscosity in our jets. We note, however, that flow focusing on the meniscus creates jets that have a characteristic size $\sim$1/3 the diameter of the rectangular (R) channels. A reduction of our calculated Reynolds numbers by a factor of three does little to stifle the apparent inertia dominance. Our experimental range in Weber number, $2-664$, indicates that for our slowest jets, surface tension plays a large role in their behavior. The slowest jets arise from R3 channels and experience rapid ligament disintegration into droplets. It is of no surprise from the dominance of inertia and surface tension that the Ohnesorge number is low for all channels. The Ohnesorge number $Oh = 0.0104, 0.0088$, and 0.0083 for (C), (R2), and (R3), respectively.

\begin{figure}
    \centering
    \includegraphics[width = 0.7\textwidth]{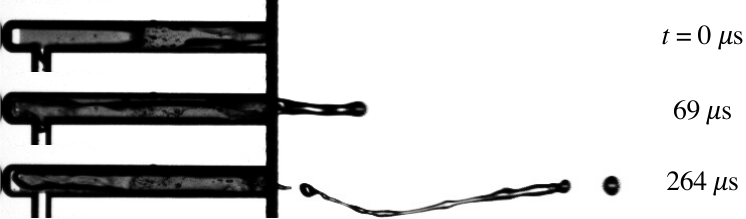}
    \caption{The jetting sequence produced by a relatively large bubble, resulting in a swaying jet tail. The panels show the moments of jet emergence (top), complete bubble collapse (middle), and when the final droplet breaks from the main jet body within the channel (bottom). (Multimedia View)}
    \label{S2}
\end{figure}

\subsection{Jet breakup and the breakup factor}
\begin{figure}[H]
    \centering
    \includegraphics[width=0.8\textwidth]{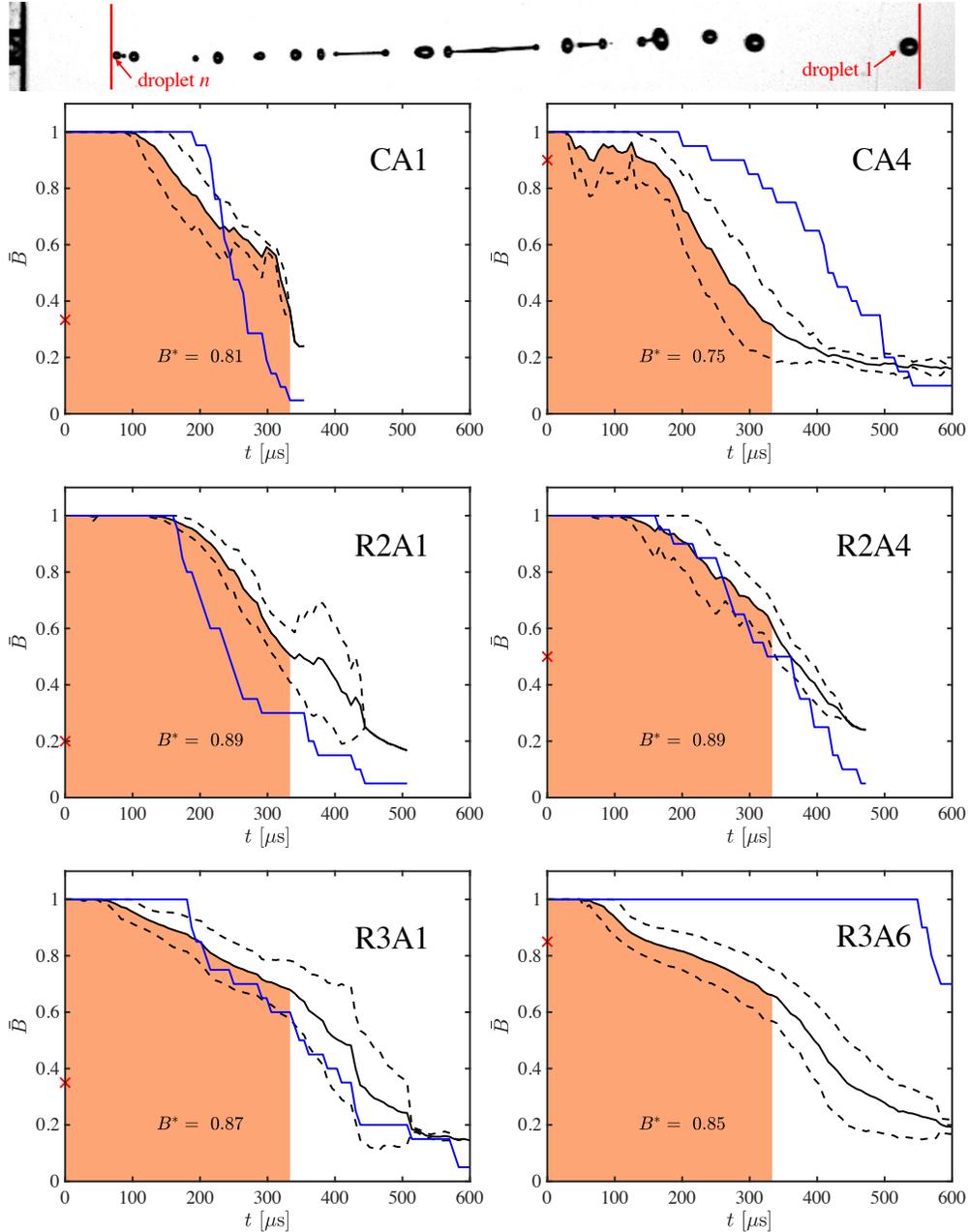}
    \caption{Breakup factor versus jetting time for featured channel configurations. Dotted lines represent standard deviation bounds, which are limited to not exceed $B=1$. Values printed beneath curves correspond to \eq{B*}. Blue curves represent the fraction of trials contributing to $\bar{B}(t)$ and the red $\times$ on the ordinates correspond to the number of trials in which leading drops leave the right-hand side of the FOV. \textbf{Top Panel:} A representative photograph of jet breakup with red lines bounding the breakup window in \eq{B}. The jet, emitted from an R2A6 channel has $U=8.64$ m/s, $Re=1350$, $We=142$, and $T=583$ $\mu$s.}
    \label{bf}
\end{figure}

 For a quantitative measure of how coherent jets break into drops throughout the jetting event, we define the `breakup factor' $B$. The breakup factor is a ratio of the total liquid parcel length along the jetting axis to the distance between the tip of the leading drop to the tail of the tailing drop. To make $B$ comparable across different channels we define the \textit{primary jetting event} from the moment the jet tip leaves the chip at $t=0$ to the moment the leading drop leaves the frame at $t=T$. The average value of $T$ for each channel $\bar{T}$ across $N$ trials is reported in \tab{params_tab}. 
 Most often jets exit the FOV at the rightmost edge (approximately 4850 $\mu$m from the channel exit) but may exit earlier through the top or bottom of the FOV.  An example of the window taken to measure $B$ is shown by the bounding red lines in the top panel of \fig{bf}. Mathematically the breakup factor is represented by, 
\begin{equation}
    B(t) = \sum_{i=1}^\text{n}D_i\frac{1}{(x(t)_{\text{c},1}+D_{1}/2)-(x(t)_{\text{c},i}-D_{i}/2)}, \label{B}
\end{equation}
where $n$ is the number of liquid parcels (either drops or ligaments) in the observation window, 
$D_i$ is the equivalent diameter of the $i$-th liquid parcel, and $x_{\text{c},i}$ is the lateral centroid location of a liquid parcel. The denominator of \eq{B} is the width of the window over which the breakup factor is measured.
A nozzle emitting a single drop, or an unbroken column of liquid, has a breakup factor of unity for all time. We average the breakup factor across $N$ videos and plot $\bar{B}$  versus jetting time $t$ in \fig{bf} for our featured channels. Since each individual jet leaves the frame at a different time, the number of trials used to calculate the $\bar{B}$ curve reduces as time progresses. The blue curves represent the fraction of trials $N$ that contribute to $\bar{B}(t)$. The red `$\times$' on the vertical axes represents the fraction of videos in which the leading drop exits the frame on the right, rather than the top or bottom. The area under the $\bar{B}(t)$ curve can be compared to an unbroken jet that maintains $B=1$ for some specified time $\tau$. Accordingly, we define this ratio as
\begin{equation}
   B^* = \frac{1}{\tau} \int_0^{\tau} \bar{B}\mathrm{d}t \label{B*}.
\end{equation}
We set $\tau=1/3$ ms such that we can compare $B^*$ values across all channel configurations; after this time $\bar{B}$ is undefined for some channels because all their respective leading drops have reached the boundary of the FOV. We report the values of $B^*$ in the plots of \fig{bf} and \tab{params_tab}. The choice of $\tau$ plays a large role in the value of $B^*$. At short times (up to $\tau \approx 0.1$ ms), $B^*$ tends toward unity, while for the circular channels $B^*(0.5~\text{ms} ) = 0.55 \pm 0.02$ except for CA1 where $B^*(0.5~\text{ms} )$ is not defined. Similarly, for the rectangular channels $B^*(0.5~\text{ms} ) = 0.72 \pm 0.02$ except for R2A4 for which $B^*(0.5~\text{ms} )$ is not defined and R3A4 for which $B^*(0.5~\text{ms} ) = 0.51$. In other words $B^*$ decreased $\approx$ 25 $\pm$ 5 \%  from $\tau = 1/3$ ms to $\tau = 1/2$ ms.  
Of our featured channels, R3A6 retains the most videos through time of any channel but has the lowest average jet velocity, a likely contributor to its propensity for producing axially focused jet trajectories. 

From the values of B*, we find that hydrophobic coatings do not promote or delay breakup in comparison to homogeneous channels. 

\subsection{Droplet size distribution}
\begin{figure}[H]
    \centering
    \includegraphics[width=0.8\textwidth]{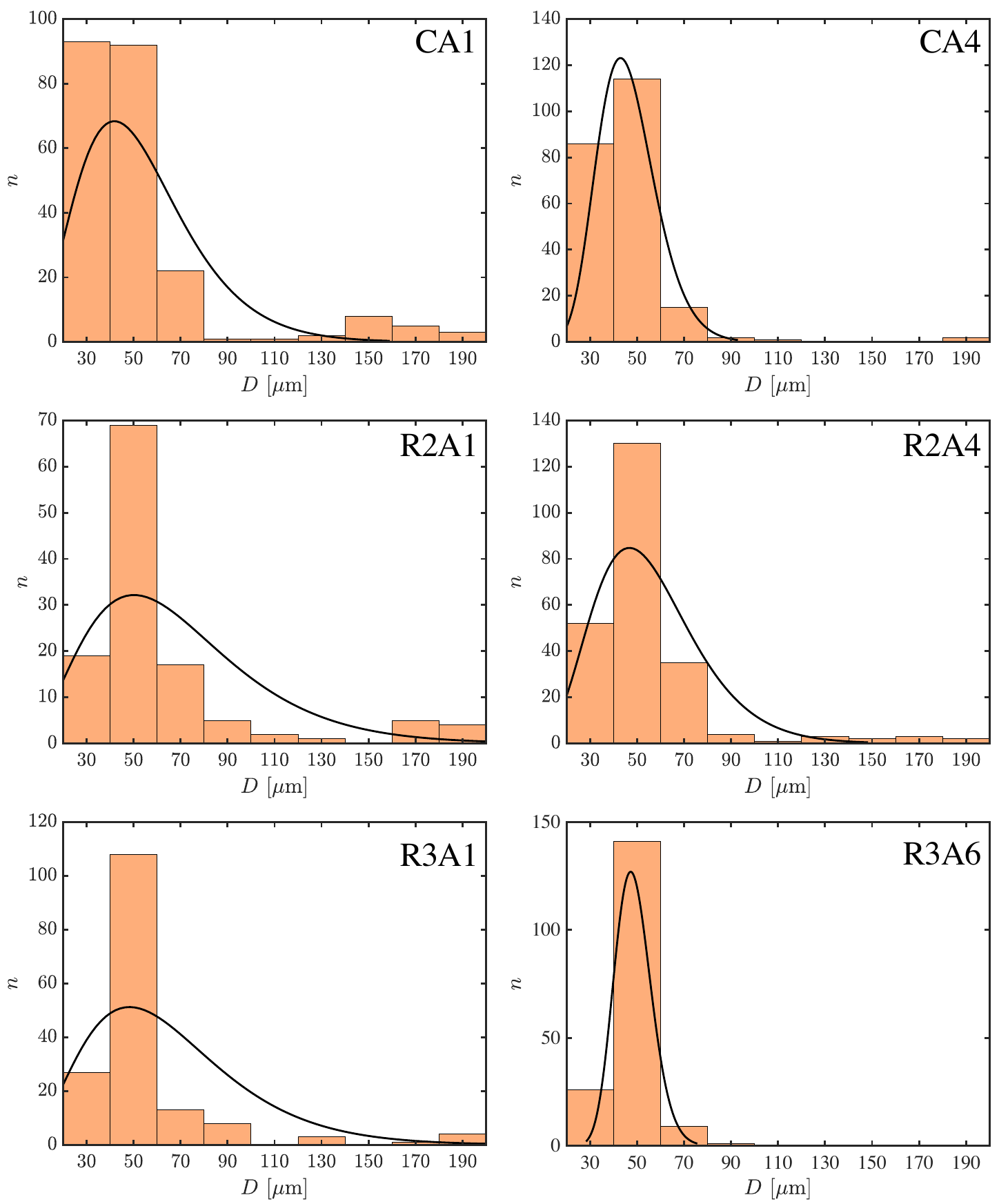}
    \caption{Drop distribution histograms at $t=T$ for featured nozzles. Histogram bin sizes are 20 $\mu$m. Solid lines represent a gamma distribution of the histogram. $\bar{T}$ is provided for all channels in \tab{params_tab}.}
    \label{drop_dist}
\end{figure}

The distribution of equivalent drop diameter at the end frame of the primary jetting event, $t=T$ is shown for featured nozzles in \fig{drop_dist}. Bin sizes are 20 $\mu$m, starting at 20 $\mu$m. Below 20 $\mu$m drops are less than three pixels across and are filtered by our binarization algorithm. The majority of drops present at $t=T$ range from 40-60 $\mu$m, a dominance that is generally enhanced by coating. 
The droplet breakup in our system is highly dependent on multiple factors and varies even for jets ejected at similar conditions. This leads to a wide size distribution that is typical for uncontrolled breakups \cite{marmottant2004fragmentation}. As done previously for these random breakup processes that involve fragmentation and coalescence \cite{marmottant2004fragmentation,keshavarz2016ligament,kooij2018determines}, we fit a gamma distribution to the drop size histogram,
\begin{equation}
    n =\frac{1}{k^\alpha \Gamma(\alpha)}D^{\alpha -1} e^{\frac{-D}{k}},
\end{equation}
where $n$ is the number of drops in a bin, $\alpha$ is the shape parameter and $k$ the scale parameter. In this case, $\alpha$ gives an idea of how `corrugated' the jet is, where $\alpha =1$ for a smooth jet and for a corrugated one $\alpha > 1$. While $k$ is directly related to the width of the distribution. We expect the shape factor $\alpha$ to be correlated with jet corrugation $B^*$, as both measure the jet roughness. However, we found that $\alpha$ and $B^*$ are not related. There are two reasons for this: i) both parameters are measured at different times, $B^*$ is measured at $\tau= 1/3$ ms whereas $\alpha$ is evaluated at $t=T$. ii) $B^*$ depends on the ratio between white pixels and black pixels in the viewing window, regardless of the jet structure. In contrast, $\alpha$ depends on how the jet breaks up, smaller and larger numbers of droplets lead to larger $\alpha$. We report $\alpha$ and $k$ for all channels in \tab{params_tab}.

\begin{scriptsize}
\begin{longtable}{|c|ccccccccccccc|} 
\caption{Jet characterization parameters. Featured channels are highlighted in \textcolor{red}{red}. $N$ indicates the number of videos. $U/\bar{U}_\text{bub}$ indicates the slope of the linear fit of $U(\bar{U}_\text{bub})$, and R$^2$, $U\sim\bar{U}_\text{bub}$ indicates the R$^2$-value representing the quality of this fit. $B^*$ indicates the breakup factor, $F$ the focusing factor. $\alpha$ and $k$ respectively indicate the shape and scale factor for the gamma distribution of the drop distribution.} \label{params_tab}
\\
\hline

\begin{tabular}[c]{@{}c@{}}Geometric\\ Configuration\end{tabular} & \multicolumn{3}{c|}{\textbf{C}}                                                                        & \multicolumn{5}{c|}{\textbf{R2}}                                                                                                                                             & \multicolumn{5}{c|}{\textbf{R3}}                                                                                                                        \\ \hline
\endfirsthead
\endhead
\begin{tabular}[c]{@{}c@{}}Coating\\ Configuration\end{tabular}   & \multicolumn{1}{c|}{\textcolor{red}{\textbf{A1}}} & \multicolumn{1}{c|}{\textbf{A2}} & \multicolumn{1}{c|}{\textcolor{red}{\textbf{A4}}} & \multicolumn{1}{c|}{\textcolor{red}{\textbf{A1}}} & \multicolumn{1}{c|}{\textbf{A2}} & \multicolumn{1}{c|}{\textcolor{red}{\textbf{A4}}} & \multicolumn{1}{c|}{\textbf{A6}} & \multicolumn{1}{c|}{\textbf{A8}} & \multicolumn{1}{c|}{\textcolor{red}{\textbf{A1}}} & \multicolumn{1}{c|}{\textbf{A2}} & \multicolumn{1}{c|}{\textbf{A4}} & \multicolumn{1}{c|}{\textcolor{red}{\textbf{A6}}} & \textbf{A8} \\ \hline

$N$ & \multicolumn{1}{c|}{21}          & \multicolumn{1}{c|}{20}          & \multicolumn{1}{c|}{20}          & \multicolumn{1}{c|}{20}          & \multicolumn{1}{c|}{20}          & \multicolumn{1}{c|}{20}          & \multicolumn{1}{c|}{42}         & \multicolumn{1}{c|}{20}          & \multicolumn{1}{c|}{20}          & \multicolumn{1}{c|}{22}          & \multicolumn{1}{c|}{20}          & \multicolumn{1}{c|}{20}         & 21
\\ \hline

$U/\bar{U}_\text{bub}$ & \multicolumn{1}{c|}{0.87}          & \multicolumn{1}{c|}{0.58}          & \multicolumn{1}{c|}{0.68}          & \multicolumn{1}{c|}{0.83}          & \multicolumn{1}{c|}{0.75}          & \multicolumn{1}{c|}{0.80}          & \multicolumn{1}{c|}{0.81}         & \multicolumn{1}{c|}{0.85}          & \multicolumn{1}{c|}{0.90}          & \multicolumn{1}{c|}{1.52}          & \multicolumn{1}{c|}{1.29}          & \multicolumn{1}{c|}{1.07}         & 1.16
\\ \hline

R$^2$,  $U\sim\bar{U}_\text{bub}$  & \multicolumn{1}{c|}{0.84}          & \multicolumn{1}{c|}{0.79}          & \multicolumn{1}{c|}{0.75}          & \multicolumn{1}{c|}{0.95}          & \multicolumn{1}{c|}{0.73}          & \multicolumn{1}{c|}{0.94}          & \multicolumn{1}{c|}{0.92}         & \multicolumn{1}{c|}{0.90}          & \multicolumn{1}{c|}{0.91}          & \multicolumn{1}{c|}{0.43}          & \multicolumn{1}{c|}{0.70}          & \multicolumn{1}{c|}{0.27}         &  0.35
\\ \hline

$Re$                                                             
& \multicolumn{1}{c|}{\begin{tabular}[c]{@{}c@{}c@{}}  2000 \\ $\pm$303 \end{tabular}}   & \multicolumn{1}{c|}{\begin{tabular}[c]{@{}c@{}c@{}}  1409 \\ $\pm$234 \end{tabular}}   & \multicolumn{1}{c|}{\begin{tabular}[c]{@{}c@{}c@{}} 1325 \\ $\pm$359 \end{tabular}}          & \multicolumn{1}{c|}{\begin{tabular}[c]{@{}c@{}c@{}} 1812 \\ $\pm$238 \end{tabular}}          & \multicolumn{1}{c|}{\begin{tabular}[c]{@{}c@{}c@{}}  1323 \\ $\pm$144 \end{tabular}}         & \multicolumn{1}{c|}{\begin{tabular}[c]{@{}c@{}c@{}} 2105 \\ $\pm$412 \end{tabular}}          & \multicolumn{1}{c|}{\begin{tabular}[c]{@{}c@{}c@{}}  1545\\ $\pm$301 \end{tabular}}         & \multicolumn{1}{c|}{\begin{tabular}[c]{@{}c@{}c@{}}   1660\\ $\pm$234 \end{tabular}}          & \multicolumn{1}{c|}{\begin{tabular}[c]{@{}c@{}c@{}}   1967 \\ $\pm$347  \end{tabular}}          & \multicolumn{1}{c|}{\begin{tabular}[c]{@{}c@{}c@{}}   892\\ $\pm$596 \end{tabular}}          & \multicolumn{1}{c|}{\begin{tabular}[c]{@{}c@{}c@{}}   1501\\ $\pm$465  \end{tabular}}          & \multicolumn{1}{c|}{\begin{tabular}[c]{@{}c@{}c@{}}   1210\\ $\pm$259 \end{tabular}}         &        \begin{tabular}[c]{@{}c@{}c@{}}   1352\\ $\pm$344 \end{tabular}                                                                                   \\   \hline

$We$                                                             & \multicolumn{1}{c|}{\begin{tabular}[c]{@{}c@{}c@{}} 444  \\ $\pm$134 \end{tabular}}   & \multicolumn{1}{c|}{\begin{tabular}[c]{@{}c@{}c@{}} 221  \\ $\pm$71 \end{tabular}}          & \multicolumn{1}{c|}{\begin{tabular}[c]{@{}c@{}c@{}}  204 \\ $\pm$116  \end{tabular}}          & \multicolumn{1}{c|}{\begin{tabular}[c]{@{}c@{}c@{}} 261\\ $\pm$67 \end{tabular}}          & \multicolumn{1}{c|}{\begin{tabular}[c]{@{}c@{}c@{}} 138 \\ $\pm$29 \end{tabular}}          & \multicolumn{1}{c|}{\begin{tabular}[c]{@{}c@{}c@{}} 359 \\ $\pm$148 \end{tabular}}          & \multicolumn{1}{c|}{\begin{tabular}[c]{@{}c@{}c@{}} 172 \\ $\pm$69 \end{tabular}}         & \multicolumn{1}{c|}{\begin{tabular}[c]{@{}c@{}c@{}} 219 \\ $\pm$64 \end{tabular}}          & \multicolumn{1}{c|}{\begin{tabular}[c]{@{}c@{}c@{}} 278\\ $\pm$99 \end{tabular}}          & \multicolumn{1}{c|}{\begin{tabular}[c]{@{}c@{}c@{}} 79\\ $\pm$70 \end{tabular}}          & \multicolumn{1}{c|}{\begin{tabular}[c]{@{}c@{}c@{}} 171\\ $\pm$124 \end{tabular}}          & \multicolumn{1}{c|}{\begin{tabular}[c]{@{}c@{}c@{}} 106\\ $\pm$41 \end{tabular}}         &         \begin{tabular}[c]{@{}c@{}c@{}} 135\\ $\pm$52 \end{tabular}                                                                                     \\   \hline

${\bar{T}}$ ($\mu$s)                              & \multicolumn{1}{c|}{\begin{tabular}[c]{@{}c@{}c@{}} 263  \\ $\pm43$ \end{tabular}}   & \multicolumn{1}{c|}{\begin{tabular}[c]{@{}c@{}c@{}} 358  \\ $\pm102$ \end{tabular}}          & \multicolumn{1}{c|}{\begin{tabular}[c]{@{}c@{}c@{}} 432 \\ $\pm121$ \end{tabular}}          & \multicolumn{1}{c|}{\begin{tabular}[c]{@{}c@{}c@{}} 365  \\ $\pm170$ \end{tabular}}          & \multicolumn{1}{c|}{\begin{tabular}[c]{@{}c@{}c@{}} 370  \\ $\pm162$ \end{tabular}}          & \multicolumn{1}{c|}{\begin{tabular}[c]{@{}c@{}c@{}} 334  \\ $\pm89$ \end{tabular}}          & \multicolumn{1}{c|}{\begin{tabular}[c]{@{}c@{}c@{}} 458  \\ $\pm180$ \end{tabular}}         & \multicolumn{1}{c|}{\begin{tabular}[c]{@{}c@{}c@{}} 427  \\ $\pm157$ \end{tabular}}          & \multicolumn{1}{c|}{\begin{tabular}[c]{@{}c@{}c@{}} 408  \\ $\pm167$ \end{tabular}}          & \multicolumn{1}{c|}{\begin{tabular}[c]{@{}c@{}c@{}} 484  \\ $\pm266$ \end{tabular}}          & \multicolumn{1}{c|}{\begin{tabular}[c]{@{}c@{}c@{}} 521  \\ $\pm184$ \end{tabular}}          & \multicolumn{1}{c|}{\begin{tabular}[c]{@{}c@{}c@{}} 563  \\ $\pm196$ \end{tabular}}         &         \begin{tabular}[c]{@{}c@{}c@{}} 491  \\ $\pm178$ \end{tabular}                                                                                        \\ \hline

\begin{tabular}[c]{@{}c@{}}${B^*}$  \\ ($\tau=1/3$ ms)  \end{tabular}                           & \multicolumn{1}{c|}{0.81}          & \multicolumn{1}{c|}{0.71}          & \multicolumn{1}{c|}{0.75}          & \multicolumn{1}{c|}{0.89}          & \multicolumn{1}{c|}{0.91}          & \multicolumn{1}{c|}{0.89}          & \multicolumn{1}{c|}{0.87}         & \multicolumn{1}{c|}{0.89}          & \multicolumn{1}{c|}{0.87}          & \multicolumn{1}{c|}{0.84}          & \multicolumn{1}{c|}{0.64}          & \multicolumn{1}{c|}{0.85}         &         0.84                                                                                        \\ \hline

\begin{tabular}[c]{@{}c@{}}shape\\parameter, $\alpha$               \end{tabular}                         & \multicolumn{1}{c|}{4.62}          & \multicolumn{1}{c|}{16.87}          & \multicolumn{1}{c|}{13.73}          & \multicolumn{1}{c|}{3.70}          & \multicolumn{1}{c|}{6.06}          & \multicolumn{1}{c|}{6.11}          & \multicolumn{1}{c|}{8.48}          & \multicolumn{1}{c|}{5.17}          & \multicolumn{1}{c|}{3.81}          & \multicolumn{1}{c|}{5.90}          & \multicolumn{1}{c|}{7.23}          & \multicolumn{1}{c|}{38.04}          & 7.95        \\ \hline

\begin{tabular}[c]{@{}c@{}}scale\\parameter, $k$           \end{tabular}                               & \multicolumn{1}{c|}{11.57}          & \multicolumn{1}{c|}{2.73}          & \multicolumn{1}{c|}{3.37}          & \multicolumn{1}{c|}{18.60}          & \multicolumn{1}{c|}{9.42}          & \multicolumn{1}{c|}{9.15}          & \multicolumn{1}{c|}{6.52}         & \multicolumn{1}{c|}{11.52}          & \multicolumn{1}{c|}{17.17}          & \multicolumn{1}{c|}{10.52}          & \multicolumn{1}{c|}{7.23}          & \multicolumn{1}{c|}{1.28}         & 7.35         \\ \hline



$F$ [$\mu$m]                                  & \multicolumn{1}{c|}{138.9}          & \multicolumn{1}{c|}{92.6}         & \multicolumn{1}{c|}{92.1}          & \multicolumn{1}{c|}{238.2}          & \multicolumn{1}{c|}{195.2}         & \multicolumn{1}{c|}{103.8}         & \multicolumn{1}{c|}{121.4}          & \multicolumn{1}{c|}{111.6}          & \multicolumn{1}{c|}{161.1}          & \multicolumn{1}{c|}{208.4}         & \multicolumn{1}{c|}{82.2}         & \multicolumn{1}{c|}{69.3}          & 
124.2          \\ \hline

\end{longtable} 
\end{scriptsize}

\subsection{Jet trajectories} \label{jet_tip_behavior}

\begin{figure}[h]
    \centering
    \includegraphics[width=\textwidth]{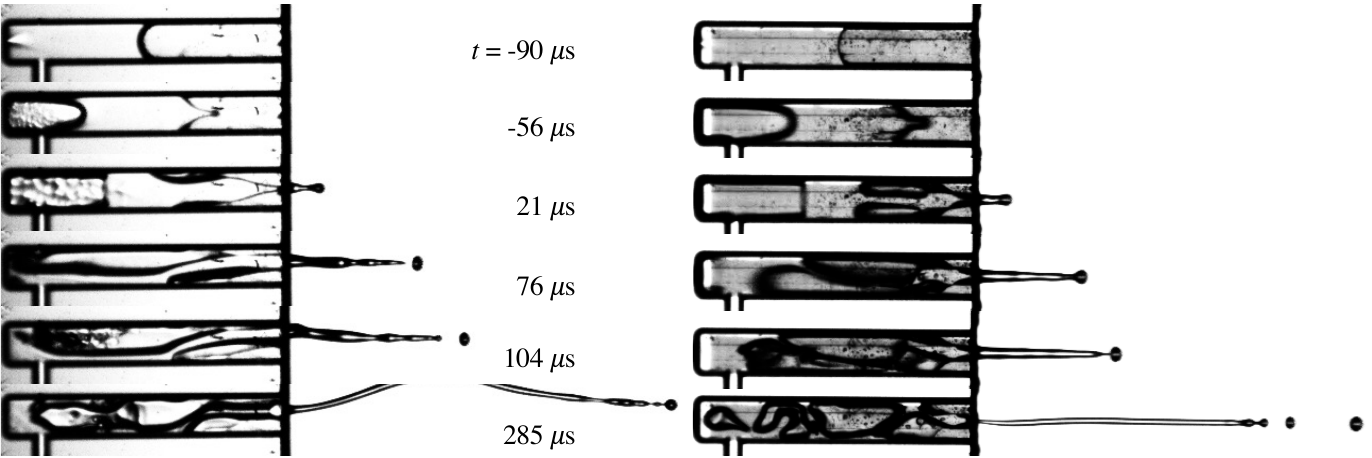}
    \caption{Time sequences of bubble expansion and partial collapse for R3A1 (left, $T=382$ $\mu$s, $U=10.96$ m/s, $Re=1920$, $We=257$) and R3A6 (right, $T=500$ $\mu$s, $U=8.36$ m/s, $Re=1470$, $We=150$) at comparable time steps following nucleation ($t=0$).}
    \label{R3_bubs}
\end{figure}

We find the shape of the static meniscus formed during channel filling to be the primary factor influencing the directional bias of jet tips. Symmetric static menisci wet the opposing walls equally to produce jet tips that exit the channel aligned with the channel centerline. Channels R3A1 and R3A6 pictured in \fig{R3_bubs} have a symmetric coating pattern when viewed from the front and thus when filled have a symmetric static menisci. From \fig{R3_bubs} the formation of self-focused jet tips is observed at $t=-56$ $\mu$s and the exit of these focused tips from the channels is seen at $t>0$ $\mu$s.

Asymmetrically coated channels form asymmetric static menisci that bias the jet tip trajectory. We present the average trajectory of the jet tip  in \fig{leading_drops}. The variance in trajectories indicates that coatings, by way of the static menisci shape, influence the direction of the average leading drop.  
Average trajectories are calculated as follows: first, the individual jet tip trajectories are obtained from the videos. Then, for all $x$-values, the average is taken of all individual jet tips. In some cases, the jet tip leaves the FOV through the top or bottom edge ($y = \pm$~250~µm or $z = \pm $~320~µm) instead of the rightmost edge (at $x$~=~4750~µm). In this case, an individual trial does not contribute to the average trajectory plot for $x$-values larger than where it left the FOV. The shaded regions in \fig{leading_drops} indicate the standard deviations from the average trajectory. Front view trajectories ($x,y$ view) are available for all channels and are shown in orange. The trajectories of the top view ($x,z$), where available, are shown in blue.

\subsubsection*{Front view $y$ trajectories}
For all channels, including those with hydrophobic coatings, their static menisci are expected to be symmetric in the $y$ direction, due to the symmetry of the coatings in $y$. Gravitational forces in our system are negligible. Therefore, we do not expect any systematic directionality in the front view (\fig{R3_bubs}). However, local surface defects can generate non-systematic exceptions because they may cause an initial asymmetric static meniscus. The defects are usually microscopic imperfections on the glass or gold surfaces formed during fabrication. If the defects are present at the exact locations where a channel is filled, the  asymmetric static meniscus that is subsequently formed can change jet trajectories. The presence of these defects and their significance can be discerned only by viewing jetting behavior after the experiment trial. We find this to be the case for the trajectory deviations in R3A8 and R3A6. Otherwise, as expected, the deviation of the jet in the vertical $y$-direction from the centerline does not show a systematic bias, with the exception of CA1. The CA1 directionality can be attributed to the upward tilt of the chip in its holder.

\subsubsection*{Top view $z$ trajectories}

\begin{figure}[H]
    \centering
    \includegraphics[width=.8\textwidth]{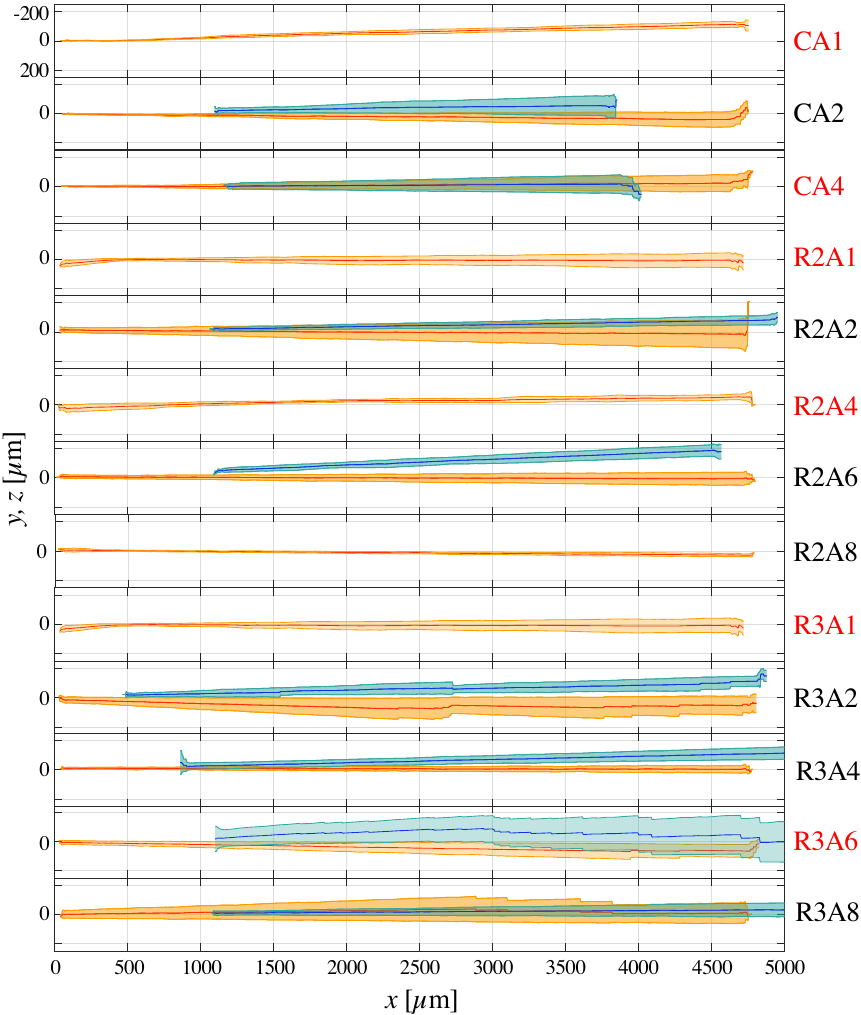}
    \caption{Average trajectory of jet tips from front (red, $y$) and top (blue, $z$) views with standard deviation shaded regions in orange and light blue, respectively. The $y,z$-dimension for each panel is $\pm 250$ $\mu$m, with an inverted axis, such that negative $y,z$-values are up. Featured channels are labeled with red text. Coatings generate no systematic bias in $y$. Coating configurations which are asymmetric about the $x-y$ plane bias jets toward $-z$ as seen from a top view.}
    \label{leading_drops}
\end{figure}

\noindent In contrast to the deviation in the $y$-direction, trajectories in $z$ show a clear bias away from the channel centerline, especially the asymmetrically coated channels (A2 and A6). Trajectories bias towards $-z$-direction, or to the right side in the channel schematics in \fig{Intro_fig}a. For the three symmetric channel configurations with top views (CA4, R3A4, R3A8), only R3A4 shows a small bias toward $-z$ (1.25\textdegree). Therefore, we can conclude that the bias of the asymmetric channels is caused by coating asymmetry and not by camera or chip misalignment.
For asymmetrically coated channels, the bias is towards the hydrophobic gold coating for A2 and towards the centered hydrophobic gold strip for A6. Of these two patterns, the A6 channels show a greater bias, and R3A6 exhibits the most extreme case of bias toward the $-z$-direction. The extreme bias of R3A6 is lost in \fig{leading_drops} because almost all the jet tips leave the FOV in the $-z$-direction and most at $x<3000$ $\mu$m, after which they no longer contribute to the average trajectory. Thereafter, the contributions of the jet tips closer to the centerline become more significant, and the average trajectory shifts back towards the centerline.

\subsubsection*{Contact line effects on jet trajectory}
\begin{figure}[H]
    \centering
    \includegraphics[width=\textwidth]{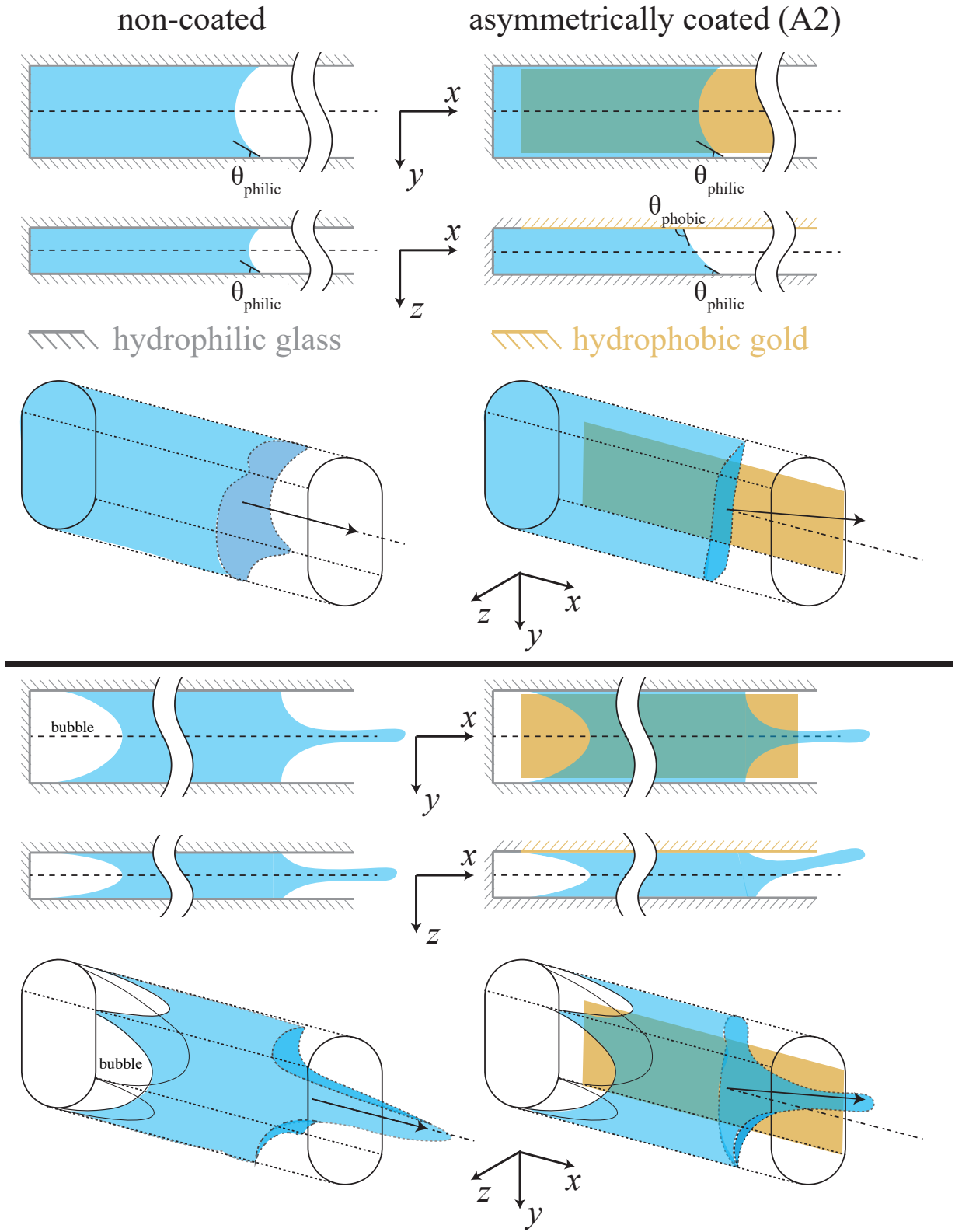}
        \caption{Front ($x,y$), top ($x,z$) and isometric ($x,y,z$) schematics of the static meniscus (upper panels) and flow-focusing effect during bubble expansion (bottom panels). A non-coated channel (A1) is shown on the left column and a coated channel (A2) on the right.}
    \label{tiltedmeniscus}
\end{figure}

Here, we explain how the contact line dynamics and the wettability of the channels affect the meniscus focusing, the tip bias and the tail movement out of axis.
As discussed previously, for asymmetric channels, the jet tips and body trajectories have a bias towards the hydrophobic coating (see \cref{front_top,leading_drops}). Initially, this bias may seem counter-intuitive, however, the static meniscus shape holds the key. For asymmetrically coated channels, the contact line at the hydrophilic walls is further advanced down the channel axis compared to the hydrophobic channel wall, as depicted in \fig{tiltedmeniscus}. Therefore, the static meniscus is slightly tilted toward the $-z$-direction, with its surface normal directed towards the hydrophobic wall. At the moment of jet formation, the tilted meniscus results in an off-axis jet. Thus meniscus shape dictates the tip direction and governs the initial stages of the jet ejection. 

\vspace{5pt}


As the jet continues to exit the channel following the leading drop, liquid is expelled and the bubble begins to retract.  The contact line on the jet side of the liquid remaining in the channel often detaches from the channel walls at different times and locations, a phenomenon we term `asymmetric contact line detachment,' which is shown for R3A1 in the image sequences of \fig{R3_bubs} (left) at $t=21$ and $76$ $\mu$s. 
Asymmetric detachment creates a deviation in the jet body away from the trajectory set by the jet tip, exhibited by the R3A1 jet at $t=104 \mu s$ in \fig{R3_bubs}. The jet body shifts toward the upper channel wall, and at $285$ $\mu$s the curved jet tail leaves the FOV. The sway does not affect the highly inertial jet tip, which continues on its course. Altogether, the jets that experience asymmetric contact line detachment have a large spread of intensities in the STDs in $y$ and a low focusing factor $F$. 

Asymmetric detachment has been observed to arise from: i) The presence of $<$20 $\mu$m sized air bubbles present on the upper or lower walls, introduced during channel filling. These bubbles burst during the rapid advancement of the contact line, contributing to detachment. ii) Contact line hysteresis due to local surface defects originating from chip fabrication.  iii) A vertical eccentricity of approximately $20-30$ $\mu$m in the location of bubble nucleation, resulting in asymmetric bubble expansion and therefore asymmetric contact-line detachment.
The presence of hydrophobic coatings can help reduce the extent of the effects of asymmetric detachment on the jet body. An example of contact line detachment stabilization can be seen \fig{R3_bubs} (right) for R3A6, where an asymmetric detachment at $t=76$ $\mu$s is not propagated to the jet body. The contrast in wetting between the hydrophilic and hydrophobic strips stabilizes the liquid bulk supplying the jet. This wetting contrast effectively creates energy barriers at the hydrophobic strips that maintain the liquid within the hydrophilic strip. Therefore the jet body continues to exit from the liquid bulk centerline, as seen at $t=104$ \& $285$ $\mu$s. 

For channel R3A8, in most cases, we observe that the jet body remains centered by adhering to the central hydrophilic strips. However, in cases with an asymmetric static meniscus, the jet tip has a bias towards the top or bottom, for which reason the jet leaves the center strip and moves toward the hydrophilic strip in the top or bottom of the channel. This deviation can be explained due to the smaller size of the centered hydrophilic strip compared to the R3A6 (89 and 98~µm for R3A8 and R3A6 respectively), as well as the smaller extent of the hydrophobic strips surrounding the hydrophilic strip (89 and 97~µm respectively). Therefore, the energy barrier for moving toward the top or bottom surfaces and wetting them is smaller, resulting in a larger fraction of the jets leaving the channel biased toward the top or bottom. For R3A4 and R3A2, the initial jet formation is centered. However, in the case of asymmetric liquid detachment, the tail of the jet sways. The sway occurs as there are no energy barriers, i.e., there is no hydrophilic strip that keeps the liquid in the center of the channel, in contrast to A6 and A8 channels.

Compared to R3 chips, the R2 chips experience a greater average bubble velocity $\bar{U}_\text{bub}$. In some cases, this larger average bubble velocity results in the formation of plug flow instead of a focused jet tip. Furthermore, the centered hydrophilic strip in R2A6 and R2A8 is smaller compared to their R3 counterparts. Therefore, the diameter of the plug is larger than the centered hydrophilic strip, resulting in the wetting of one or both hydrophobic strips. This means that the  contact line is no longer contained between the interface of the hydrophilic-hydrophobic strips and the energy barriers are overcome by the initial inertia of the system. Thus, the jets are not centered and can exit along the top or bottom of the channel.
For the circular channels, due to their smaller cross-sectional area (less than half that of R2 channels), average bubble velocities are larger than for rectangular channels, resulting in plug flow in all cases. The jet diameter is of similar size to the channel (100 $\mu$m), therefore the contact lines slide along the channel walls to the channel exit.  Because the initial cavitation bubble is of the same size as the channel length, all liquid is expelled and there is no receding contact line. Therefore, there is little sway of the jet body and tail. 





\subsection{Spatiotemporal diagrams in $x$}

The jets produced from our system behave like a high momentum fluid ligament, with pinch-off occurring as the jet travels forward \cite{villermaux2020fragmentation,eggers2008physics}. The liquid remaining in the channel acts as a `reservoir' that feeds the ligament with the expansion of the cavitation bubble. With the bubble collapse and the ejection of the remaining liquid, the ligament pinches off from the reservoir and breaks up into a string of droplets as shown in \fig{Intro_fig}c,d. A convenient way to visualize the dynamical behavior of this ligament is a spatiotemporal diagram (STD) or kymograph, which shows the evolution of the jets in a single space dimension and time. In the STDs, $x$ is the coordinate parallel to the channel's long axis, and $t$ is the perpendicular coordinate. For convenience, the edge of the channel (or nozzle) from which the liquid emerges is set as $x=0$; the time at which this occurs corresponds to $t=0$. In the STDs, therefore, we visualize the jet as it emerges from the nozzle and travels in time through the field-of-view (FOV) to the right of the nozzle. 

\begin{figure}[H]
    \centering
    \includegraphics[width=0.96\textwidth]{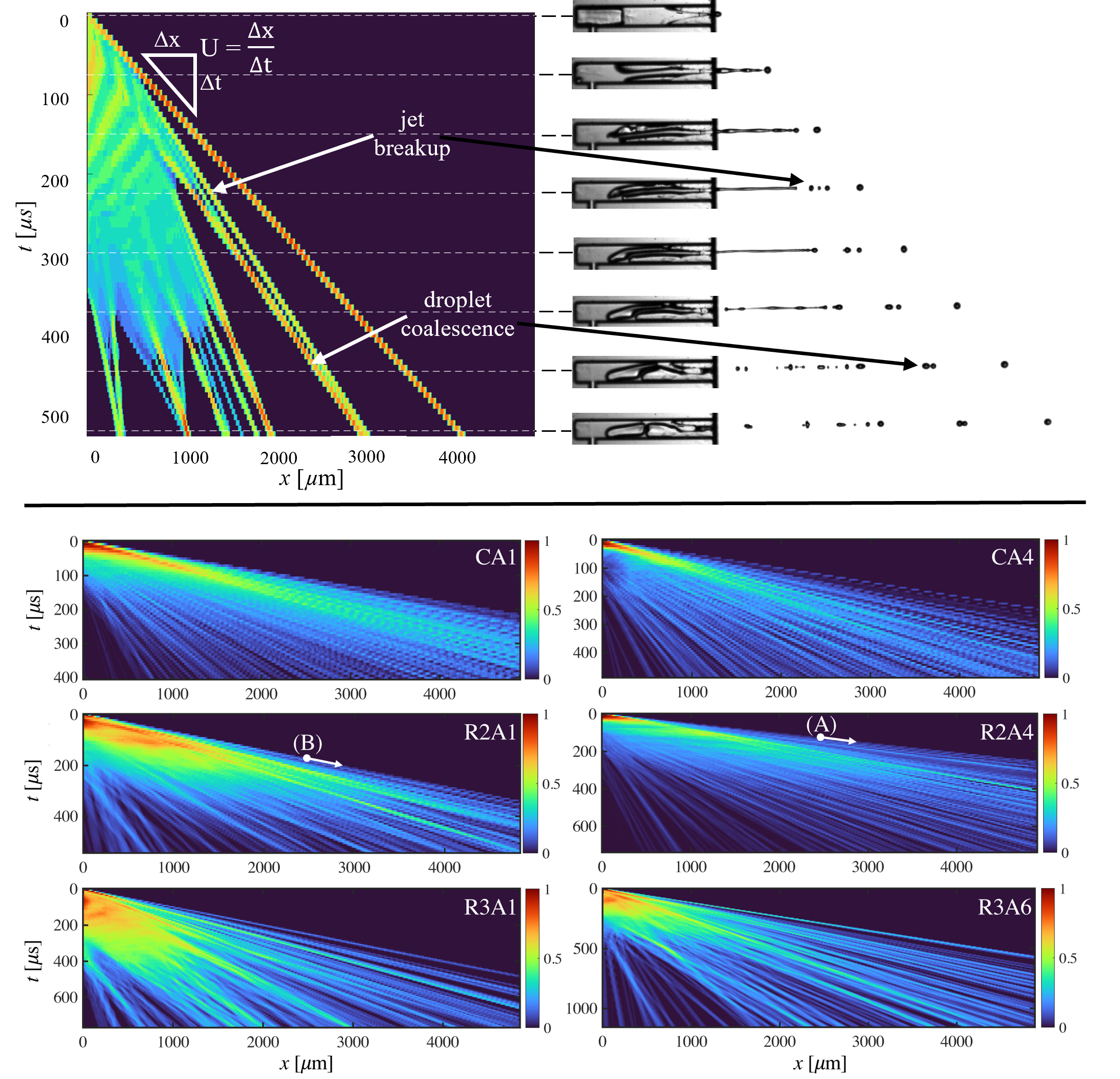}
    \caption{Spatiotemporal diagrams (STDs) in $x$. \textbf{Top:} A spatiotemporal diagram for a single video with key attribute labels. Video snapshots correspond to times indicated in the STD. \textbf{Bottom:} Aggregated $x$-$t$ STDs for featured channels. The number of trials $N$ comprising each aggregate is given in Table 2. Typically, $N=20$. Labels (A) and (B) denote trajectories of the fastest leading drops.}
    \label{std_x}
\end{figure}

STDs are created from the binarized video frames of a jetting event. Every binarized frame consists of the liquid (ligament or drops) in white against a black background. An STD created for a single video is shown adjacent to representative frames in \fig{std_x} (top).  For the STD in $x$, the binary matrix for a frame is summed along each column, resulting in a row vector with a range of `intensities'. Row vectors for each frame are stacked onto one another to form the STD in $x$, which is an $i\times j$ matrix where $i$ is the number of frames and $j$ is the number of $x$-pixels in the FOV. Liquid parcels which are longer in the direction of travel create a larger footprint in $x$ (spanning more pixels). Long unbroken lengths generally indicate ligaments, while individual lines indicate the motion of droplets. The height normal to the jetting axis (in $y$) of a liquid parcel, drop or ligament, for a single frame is represented by intensity values. Breakup is indicated by the splitting of a line and the velocity of a liquid parcel is given by the inverse of the slope of its line in the STD. Aggregated STDs in $x$ are shown for our featured channels in \fig{std_x} (bottom) which show the average axial behavior of the channel. Aggregates are formed by combining individual STDs for every trial (usually twenty) for a given channel. Individual STDs are truncated to the shortest captured video in time and averaged by the number of videos (\tab{params_tab}), then normalized by the maximum intensity to create the aggregated STDs for that channel configuration, see \fig{std_x} (bottom). Therefore all STDs in \fig{std_x} (bottom) have an intensity range from 0 to 1. 

In the aggregated STDs in \fig{std_x}, trajectories of individual drops remain distinguishable. The leading drops are found at the top surface of the wedge-like spray emanating from the origin. A shallower slope indicates a faster drop. For example, the fastest drop in R2A4 is faster than the fastest drop in R2A1, labeled by (A) and (B) in \fig{std_x}, respectively; a fact likewise confirmed by \fig{bubs}.
Jets that break up with greater consistency or repeatability across trials create aggregated STDs with fewer lines or tracks, with each track having a higher intensity due to repeated superposition of individual jetting events. R3A1 and R3A6 channels, for example, break up more repeatably across trials than the other featured channels.

\subsection{Spatiotemporal diagrams in $y$}
\begin{figure}[H]
    \centering
    \includegraphics[width=\textwidth]{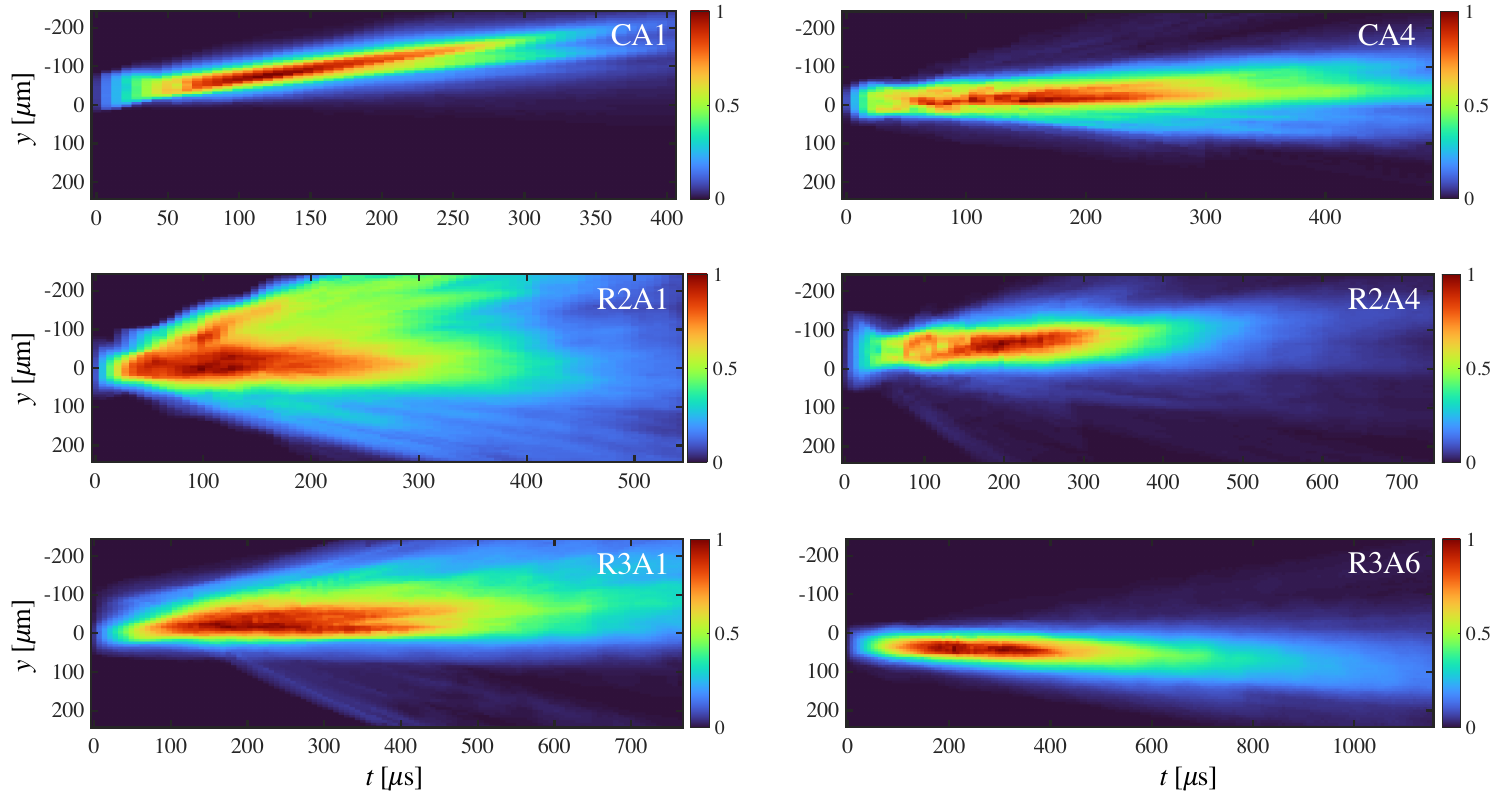}
    \caption{Aggregated $y$-$t$ spatiotemporal diagrams for featured channels.}
    \label{std_y}
\end{figure}

Another convenient means of visualizing the trajectories of jets is an STD in the $y$-direction. STDs in $y$ are created following the same approach as STDs in $x$, but the binary matrix in each frame is now summed along each row and reduced to a column vector. Column vectors are stacked in time to form the STD in $y$, which is a $k\times i$ matrix where $k$ is the number of $y$-pixels in the FOV and $i$ is the number of frames. Aggregated STDs in $y$ are shown for our featured channels in \fig{std_y}. The $y$ origin runs along the nozzle centerline. Dimensional or breakup information of the jet is not obtained from the STD in $y$; the plot instead gives the tendency to find a liquid parcel at a given $y$ location in the FOV. Moreover, individual trajectories cannot be discerned, but net deviations can be visualized. 
Initially, when jets emerge from the nozzle, intensity values gradually increase as the liquid is drained from the channel and into the ligament. The intensity values reduce as the liquid parcels either exit the FOV or deviate from the centerline. The intensity reduction in the case of centerline deviation is also complemented by an increase in non-zero pixel rows in the column vector of each frame. 

Individual plots are then aggregated by the same method as STDs in $x$. In the aggregated $y$-$t$ STDs, focused and repeatable jets are those where the trajectories are narrow and have greater intensities as a result of superimposition. Such is the case when trailing drops follow the leading drop, and an overall jet trajectory can be distinguished. The STD can either be centered or skewed in one direction if there is a preferential jetting direction. In cases where there is a different motion of the tail with respect to the jet tip, there is a spread in the STDs in $y$. In \ref{std_y}, the most repeatable jets are emitted from CA1 and R3A6. R2A1 has a large spread and therefore lower repeatability. The cause for the upward trajectory of jets produced by CA1 is unknown but again is likely the result of a slight upward tilt of the chip in its holder (approximately 0.5\textdegree).

\subsection{Focusing factor} \label{focusingfactor}
\begin{figure}[H]
    \centering
    \includegraphics[width=0.8\textwidth]{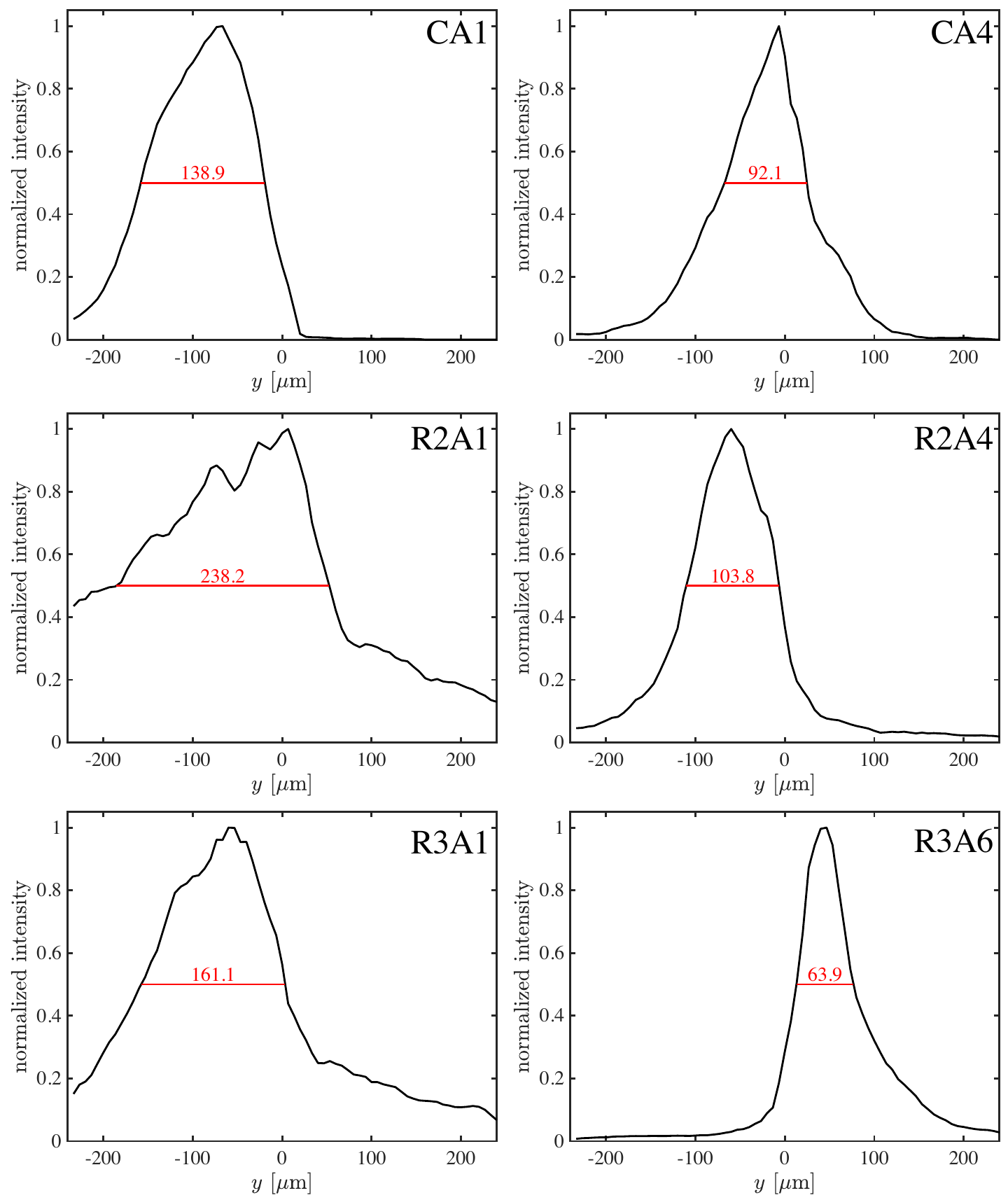}
    \caption{Maximum normalized intensity derived from STDs in $y$, across jetting time $t$ for all featured channels. Negative values indicate the top of the FOV (see \fig{introfig})}. The red lines and values report the focusing factor $F$ derived from the full width at half maximum of the normalized intensity curves. 
    \label{focus}
\end{figure}

Each column in \fig{std_y}, a snapshot in time, has a corresponding intensity curve across $y$. If time is collapsed and we take the maximum intensity value at each $y$ position, we render a single curve that represents the STD, as shown in \fig{focus}. We may quantify the aggregate focus of a jet by the focusing factor $F$, defined as the full width at half maximum (FWHM) of the intensity curves in \fig{focus}. Lower values of $F$ correspond to more focused jets. We denote $F$, which has units of $\mu$m, as a red line in \fig{focus}, and report $F$ for all channels in \tab{params_tab}. The location of the peaks in the FWHM curves, and their skewness in one direction indicate a preferential jetting direction for the chip configuration (similar to the STDs in $y$); jets that emerge and travel at an angle from the chip, have peaks at an offset from 0, a trait that is observed in almost all chips. A slight tilt in the positioning of the CA1 chip gives its jets an artificially high $F$. Otherwise, the shape of intensity curves and $F$ values correspond well to the diffusive nature of jets shown in \fig{std_y}. All coated channels improve the focusing factor $F$ over their uncoated (A1) counterparts, save R3A2. Overall, the most focused jets are produced by R3A4 and R3A6, indicating that the tallest channel (R3), inherently produces the most repeatable jets. In the upcoming subsections, we discuss in detail, the behavior of the jet tip and body and find the underlying mechanisms influencing their behavior. 

\begin{figure}[H]
    \centering
    \includegraphics[width=0.7\textwidth]{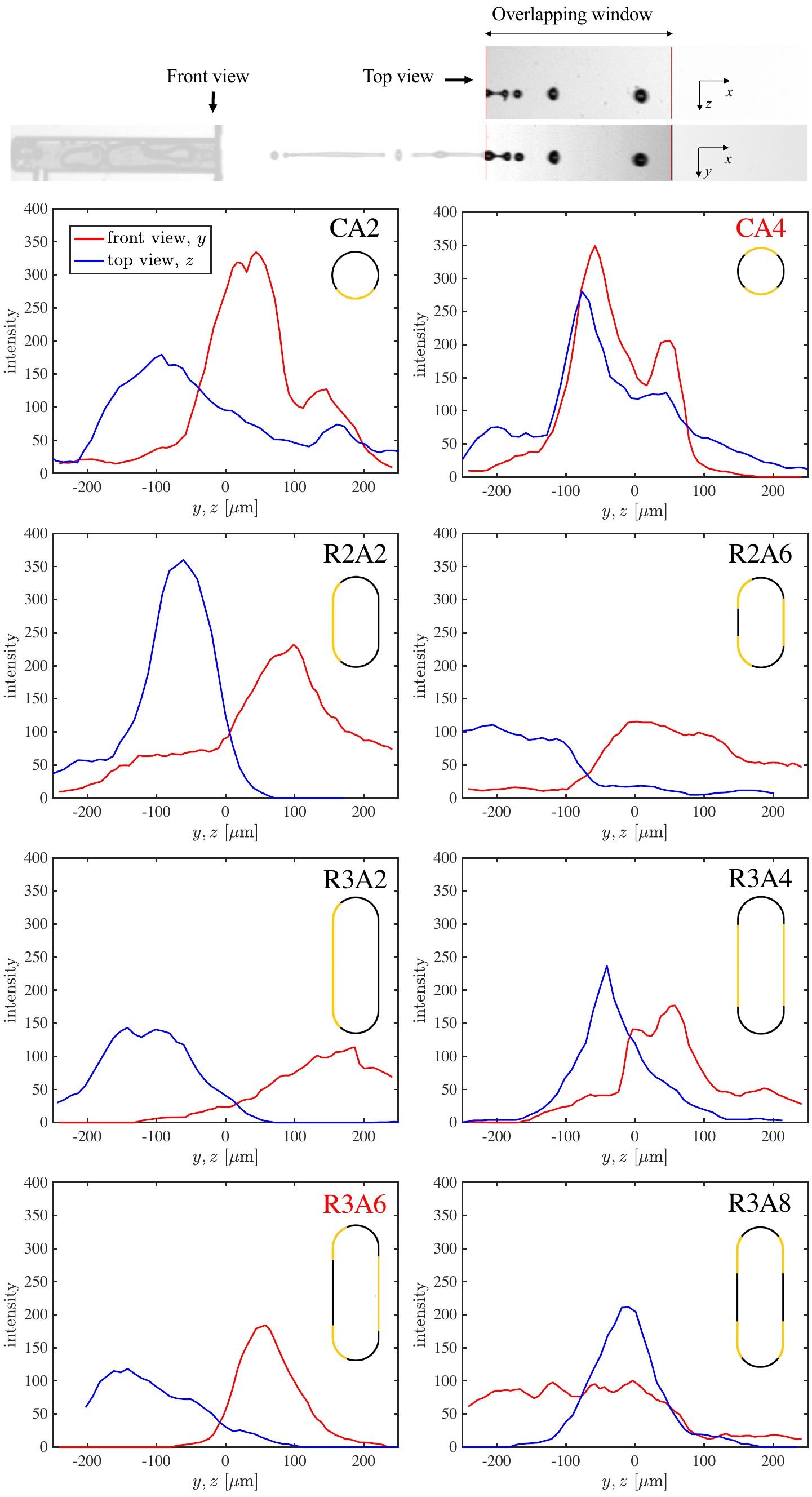}
    \caption{Maximum intensity derived from partial-view STDs in $y$ (red) and $z$ (blue), across jetting time $t$. Negative values indicate the top of the FOV. Red channel labels indicate featured channels. \textbf{Top panel}: Diagram denoting the fields of analysis in both camera views for the partial-view STDs.}
    \label{front_top}
\end{figure}

Since the effect of the bias in \fig{leading_drops} is lost, we again turn to outline curves discussed above to make a comparison with the front view trajectories. To be able to compare front and top views meaningfully, we focus on an area $1575\times482$ $\mu$m, $\approx$ 1/3rd of the FOV presented in the previous sections, located 2300 $\mu$m away from the chip edge (\fig{front_top}, top). The choice of this focus area is due to our inability to visualize the channel exit from the top. The corresponding outline curves are calculated with the same procedure as in Section \ref{focusingfactor} and are presented in \fig{front_top}. Here, as in Section \ref{focusingfactor}, the deviation of the peak from the center of these curves represents a bias in jetting direction. However, in contrast to curves in \fig{focus}, the intensity values are not normalized. The intensity values correlate to the amount of fluid passing through the window, and therefore the difference between these values allows us to compare the extent of out-of-axis behavior between both views.

We find that for the tall R3 channels, the jet body follows the directional bias (toward the top in the front view and toward the left in the top view) set by the jet tip in all cases (\fig{front_top}). For the asymmetric channels R3A2 and R3A6, the bias towards the $-z$ direction is marked by the location of the maximum values at $\approx -100$ $\mu$m for both channels. The intensity curve derived from the top camera view is lowest for R3A6 among other R3 channels, indicating that most of the jet has left the FOV before entering this window. In contrast, the symmetric channels R3A4 and R3A8 have sharp distinct peaks near the center (deviation $<50 \mu m$), with greyscale intensities $>200$, again supporting that symmetric coatings do not bias the jet trajectory. 
For the lower aspect ratio R2 channels, we also find a bias of the jet body towards the $-z$ direction for both asymmetrically coated channels. For R2A2 channel, bias in $-z$ in the top view is denoted by the presence of a sharp peak at $\approx -80$ $\mu$m. For R2A6, this $-z$ bias is seen as well; the top outline curve looks similar to that of R3A6, following an extreme out-of-axis behavior. The maximum value of the front outline curves is at $y = 0$, with a wide distribution over $y$. This distribution is owed to plug flow in R2A6 instead of a focused jet tip, which makes the jet exit from the top or bottom of the channel. 

For the circular channels, we find the bias towards the $-z$ direction expected for the asymmetrically coated CA2, and the centerline trajectory expected for the symmetrically coated CA4. The gradual decrease in intensities as we go from the circular and rectangular R2 to the rectangular R3 is owing to the greater tendency to form plug flows in the smaller channels, which is not so in the larger channels. The more complete emptying of the smaller channels due to the plug flow induced in the channel following the initial cavitation event gives rise to their greater intensity values.

\section{Concluding remarks}\label{conclusion}
In this experimental study, we present jet behaviors observed from micro-channels of three geometries with up to five coating configurations each. Channels are coated with alternating hydrophobic and hydrophilic bands along their periphery. Jets are generated by laser-induced thermocavitation and the channels are initially partially-filled such that the advancing meniscus is kinematically focused. Modifications to the rapidly accelerated meniscus by the different coatings influences the jet breakup, the resulting drop size distribution, the trajectory of the jet tip, and the consistency of jet characteristics across trials.  
Our findings agree with previous studies that the jet velocity $U$ has a linear relationship with the bubble growth velocity $\bar{U}_\text{bub}$, $U\sim\bar{U}_\text{bub}$, as shown in \fig{bubs}. No effect of the hydrophobic coatings is observed for either the circular or the rectangular cross-sectioned R2 channels. In contrast, for the higher aspect ratio R3 coated channels the ratio of jet to average bubble velocity $U/\bar{U}_\text{bub}$ increases compared to the uncoated channels, indicating less hydrodynamic resistance to the rapid thermocavitation event. 

We assessed how the coatings and their wettability influence the initial meniscus shape and contact line dynamics. These two factors are critical for understanding the jet tip direction and the jet body behavior. Asymmetrically coated channels produce an off-axis jet tip trajectory with a clear bias towards the hydrophobic channel wall. Although we could not image the meniscus from the top, we suggest that the asymmetrically shaped meniscus results in the observed flow focusing toward the hydrophobic wall. Furthermore, rectangular channels with a hydrophilic strip in the middle such as R3A6, reduce the out-of-axis trajectory. This is due to the hydrophobic strips geometrically delimiting the flow of the jet in the middle of the channel. The effect of the energy barriers is reduced for circular channels and R2 channels due to their tendencies to produce plug-like jets; the emitted jets wet the whole perimeter of the channel and are wider than the hydrophilic strips, resulting in low flow focusing.

For the analysis of the jet dynamics in time and space, we have developed a spatiotemporal diagram (STD) representation, which can be generated in both the $x$ and $y$ directions. STDs in $x$ give information about the jet breakup, coalescence, and the trajectory of individual drops. STDs in $y$ give information about the assymetric direction of the entire jet, tip and body. By extracting the maximum intensity of each time in the $y$-$t$ STDs we can extract profiles that show concisely the jet bias off the centerline and the focusing factor $F$. 





We avoid referring to any one channel as superior. The jetting characteristics from any particular channel may well be optimally suited to a particular application. For example, needle-free dermal injections will work best with jets that remain coherent over greater distances and exhibit limited off-axis behavior such as tail sway. Jets aimed at uniformly coating surfaces may work best with tails that deviate from the trajectory of their leading drops. The exquisite tunability of the present system through variation of geometry, heterogeneous surface chemistry, laser properties, and more pave a bright future for its adaption to a wide range of applications.

\section*{Acknowledgements}
We would like to thank the National Science Foundation CBET-1941341 and the European Research Council (ERC) under the European Union Horizon 2020 Research and Innovation Programme (grant agreement no. 851630) for support, McKenna E.M. Goss for text edits. Furthermore, we would like to thank Stefan Schlautmann for the fabrication of the microfluidic chips.

\section*{Data access}
Raw experimental videos and data are publicly available in perpetuity via OneDrive. Interested parties should contact the corresponding author for access.

\clearpage

\bibliographystyle{royal}
\bibliography{Grants_bib}

\end{document}